\documentclass[extra, referee]{gji}
\usepackage{timet,color}
\usepackage[colorlinks=true, linkcolor=blue, citecolor=blue, urlcolor=blue]{hyperref}
\usepackage{graphicx}
\usepackage{amsmath}
\usepackage{algorithm}
\usepackage{algorithmic}
\usepackage{float}
\usepackage{booktabs}

\usepackage{ulem}
\usepackage{xcolor}
\usepackage{lineno}
\usepackage{caption}
\usepackage[most]{tcolorbox}
\usepackage{hyperref}

\colorlet{delete}{blue}
\colorlet{new}{red}

\title[2.5D\ Transformer Interpolation]
  {2.5D Transformer: An Efficient 3D Seismic Interpolation Method without Full 3D Training}

\author[Changxin Wei et al.]
  {Changxin Wei$^1$, Xintong Dong$^1$, Xinyang Wang$^2$ \\
  $^1$ State Key Laboratory of Deep Earth Exploration and Imaging, College of Instrumentation and \\
  Electrical Engineering, Jilin University, Changchun \emph{130026}, China. E-mail: \href{dxt@jlu.edu.cn}{dxt@jlu.edu.cn}\\
  $^2$ College of Electric Engineering, Department of Communication Engineering, Key Laboratory of \\
  Modern Power System Simulation and Control and Renewable Energy Technology (Ministry of Education), \\
  Northeast Electric Power University, Jilin \emph{132012}, China
  }

\begin{document}

%\pagewiselinenumbers
%\linenumbers

\label{firstpage}

\maketitle

\begin{summary}
 Transformer has emerged as a powerful deep-learning technique for two-dimensional (2D) seismic data interpolation, owing to its global modeling ability. However, its core operation introduces heavy computational burden due to the quadratic complexity, hindering its further application to higher-dimensional data. To achieve Transformer-based three-dimensional (3D) seismic interpolation, we propose a 2.5-dimensional Transformer network (T-2.5D) that adopts a cross-dimensional transfer learning (TL) strategy, so as to adapt the 2D Transformer encoders to 3D seismic data. The proposed T-2.5D is mainly composed of 2D Transformer encoders and 3D seismic dimension adapters (SDAs). Each 3D SDA is placed before a Transformer encoder to learn spatial correlation information across seismic lines. The proposed cross-dimensional TL strategy comprises two stages: 2D pre-training and 3D fine-tuning. In the first stage, we optimize the 2D Transformer encoders using a large amount of 2D data patches. In the second stage, we freeze the 2D Transformer encoders and fine-tune the 3D SDAs using limited 3D data volumes. Extensive experiments on multiple datasets are conducted to assess the effectiveness and efficiency of T-2.5D. Experimental results demonstrate that the proposed method achieves comparable performance to that of full 3D Transformer at a significantly low cost.
\end{summary}

\begin{keywords}
 Machine Learning, Image processing, Neural networks, fuzzy logic
\end{keywords}

\section{Introduction}
Interpolation operation is a crucial step for improving the quality of seismic data, especially when facing geometries with large spatial sampling intervals, and recovering missing parts caused by topographical constraints, limited budgets, or receiver malfunctions. By reconstructing missing information and compensating for sparse geometries, effective interpolation can enhance the completeness and continuity of events, facilitating the subsequent inversion and high-resolution imaging \citep{chen2024low, cheng2025generative}.

To obtain the desired dense and complete seismic data, various conventional interpolation methods have been proposed over the past few decades \citep{spitz1991seismic, trad2002accurate, oropeza2011simultaneous}. These methods are theory-driven, relying on mathematical models or analytical formulations to describe the underlying physical processes. Generally, they can be divided into four categories: wave-equation-based methods, prediction filtering methods, low-rank methods, and sparse-representation methods.

The wave-equation-based methods utilize subsurface velocity models to simulate the propagation of seismic wavefields, so as to reconstruct incomplete data \citep{ronen1987wave}. \cite{ronen1987wave} proposed a trace-interpolation method based on the wave equation and prior assumption of a smooth spatial spectrum, and experimental results have demonstrated its effectiveness on both synthetic and field data. The offset-continuation differential equation-based method proposed by \cite{fomel2003seismic} showed good interpolation performance even in several structurally complex situations. However, the wave-equation based methods often require a relatively accurate velocity model as an important prerequisite, which is often unavailable in numerous real-world problems. Moreover, the heavy computation burden seriously limits their further applications.

The prediction filtering methods exploit the correlation of events in different domains to perform the interpolation operation. When applying these methods, the interpolation task is taken as a least-squares linear inverse problem, and interpolated results are generated by minimizing the misfit between the predicted and input data \citep{spitz1991seismic, liu2022bseismic}. \cite{spitz1991seismic} proposed a multichannel and model-free interpolation method for missing traces in f-x domain, exhibiting good interpolation performance in both pre- and post-stack examples. To further improve the interpolation performance, \cite{naghizadeh2007multistep} proposed a multistep autoregressive algorithm-based prediction-error method combined with Fourier-based methods to reconstruct seismic data from low to all frequencies. An important assumption of these prediction filtering methods is that events are required to be linear. However, this is usually not satisfied in real conditions, resulting in serious interpolation performance degradation.

The low-rank methods assume that the complete seismic data in a specific arrangement is of low rank at a given frequency component \citep{trickett2010, gao2013fast}, and missing traces will significantly increase its rank. Therefore, the interpolation tasks can be accomplished by reducing the rank to an optimal one. Typical low-rank methods include singular spectrum analysis \citep[SSA;][]{vautard1989singular, oropeza2011simultaneous, carozzi2021interpolated}, matrix completion \citep{ma2013three, yang2013seismic, kumar2015efficient}, principal component analysis \citep{WOLD198737, huang2016improved, wu2023high}, and Cadzow filtering \citep{cadzow2002signal, gao2013fast, naghizadeh2013multidimensional, huang2020dealiased}. \cite{oropeza2011simultaneous} proposed a simultaneous denoising and reconstruction method for seismic data based on multichannel SSA (MSSA), which resembles seismic data interpolation with the method of projection onto convex sets (POCS). Moreover, a randomized singular value decomposition is adopted to accelerate its rank reduction stage. \cite{ma2013three} enhanced the matrix completion using a designed texture-patch transformation, exhibiting superior performance to traditional POCS method. However, in these low-rank methods, the determination of optimal rank is an intricate problem, and an improper one will have a negative impact on the interpolated results \citep{ma2013three, cheng2023seismic}.

The sparse-representation methods interpolate seismic records by representing them into sparse domains. The useful signals and the missing data are treated as large- and small-amplitude coefficients, respectively. Preserving the large-amplitude coefficients and eliminating the small-amplitude ones helps to extract the useful signal components from the incomplete data \citep{chen2019interpolation}. The missing signal components are then recovered through an inverse sparse transform. These methods can be mainly categorized into mathematical transform-based \citep{trad2002accurate, yu2007wavelet, fomel2010seislet, naghizadeh2010beyond} and dictionary learning-based \citep{liang2014seismic, yu2015interpolation, wang2020fast} methods. \cite{trad2002accurate} proposed a high-resolution time-variant Radon transform (RT)-based interpolation method. Hyperbolic and elliptical RTs are implemented to perform accurate interpolation and attenuate sampling artifacts in poorly sampled common-midpoint gathers. Based on the good sparse representation ability of the shearlet transform, \cite{liu2018crossline} proposed a multi-component crossline seismic data reconstruction method based on sparse shearlet constraint inversion, achieving better results than traditional wavelet, curvelet, and shearlet methods in several extremely sparse sampling cases. However, most of the above mathematical transform-based methods are based on fixed bases \citep{liang2014seismic}. To learn basis functions adaptively, experts have developed dictionary learning-based interpolation methods. \cite{liang2014seismic} attempted to restore decimated seismic data using data-driven tight frame (DDTF) first developed by \cite{cai2014data}. The DDTF can adaptively learn from the data itself, and provides a sparser representation for the data in turn. \cite{yu2015interpolation} extended the DDTF to high-dimensional versions and achieves the simultaneous denoising and interpolation of 3D and 5D seismic data. Nevertheless, these sparse-representation methods are hampered by inherent limitations, despite their excellent anti-aliasing and interpolation performance. First, the sparsity of data is an essential assumption. Second, the choice of parameters plays a crucial role. For instance, for the curvelet transform, inaccuracies in the estimation of the mask function can directly affect interpolation performance; for the seislet transform, inaccurate estimation of the local event slope tends to deteriorate the fidelity of seismic data reconstruction. \citep{wang2019deep}.

Generally, the successful applications of these conventional methods must satisfy a certain number of prior assumptions, including an accurate subsurface velocity model, the linearity of events, the low-rank structures of seismic data, and the sparsity of signals \citep{jia2017what, wang2019deep, dong2025deep}. However, these assumptions reflect our limited knowledge of full natural systems and may not always conform to real geological conditions, thereby restricting the effectiveness and adaptability of these theory-driven interpolation approaches \citep{jia2017what, cheng2024meta}. These theory-driven methods also involve complex and sophisticated operations for parameter adjustments, which require substantial expert knowledge, hands-on experience, and considerable manual effort, making them labor-intensive and time-consuming in practical applications. Additionally, the heavy computational cost that occasionally occurs is also a major hurdle, especially when handling with large-scale and high-dimensional seismic datasets \citep{wang2019deep, dong2022seismic}.

In recent years, deep learning \citep[DL;][]{lecun2015deep} has attracted much attention. This kind of data-driven technique has the ability to learn from data itself and process large amounts of data more efficiently. These characteristics enable DL-based methods to perform well without prior assumptions and handcrafted parameters, so as to automatically process large-scale datasets. At the early stage, convolutional neural networks (CNNs) have been the primary focus. Typical CNN-based architectures, such as U-Net \citep{ronneberger2015unet, park2019reconstruction, park2021method, fang2021seismic}, Res-Net \citep{He2016CVPR, wang2018seismic, wang2019deep, liu2022aseismic}, and generative adversarial network \citep[GAN;][]{goodfellow2014generative, siahkoohi2018seismic, oliveira2018interpolating}, have shown great promise for the seismic data interpolation. Various improvements applied to CNNs have also contributed to further enhance the interpolation performance, including the attention-enhanced CNNs \citep{yu2021attention}, depthwise separable CNNs \citep{jin2023depthwise}, and multi-scale CNNs \citep{cheng2023seismic, dong2024can, dong2024seismic}. As researchers have gained deeper insights into DL-based interpolation methods, the local perception has emerged as a major bottleneck of CNNs, which hinders the further enhancement of the interpolation performance using DL. To address this issue, experts have turned to Transformer \citep{vaswani2017attention} owing to its strong ability to extract global contextual information using the self-attention mechanism. \cite{cheng2024seismic} proposed a U-shaped Swin-Transformer-based interpolation network, which is an encoder-decoder structure and can effectively reconstruct the consecutively missing traces in distributed acoustic sensing-vertical seismic profiling data. By integrating the global and local features, \cite{gao2024swin} proposed a Swin-Transformer convolutional residual network for the simultaneous denoising and interpolation of seismic data, gaining better visual performance and quantitative indices than conventional methods and CNN-based methods. A dense double branch attention Transformer (D2AT) proposed by \cite{dong2025deep} can effectively reconstruct the consecutively missing traces. The core module of D2AT is a global feature unit composed of six Swin-Transformer blocks distributed in two scales, and this method has shown great interpolation performance and generalization in both pre-stack and post-stack datasets.

Although these DL-based methods have achieved superior performance over conventional methods, they primarily focus on two-dimensional (2D) interpolation and overlook the spatial structure correlation that could be exploited in a 3D interpolation workflow \citep{wang2019deep, liu2022aseismic, dong2025deep}. However, 3D feature extraction operations usually introduce significantly greater computational burdens for DL-based interpolations. This phenomenon is particularly prominent when applying Transformer-based architectures, whose computational complexity scales quadratically with the size of input data \citep{vyas2020fast}. Moreover, extending 2D networks to 3D versions will significantly increase the number of trainable parameters \citep{ye20193d}. Several scholars have made attempts on 3D seismic data interpolation based on 3D CNNs in recent years \citep{qian2021dtae, chen2023projection, saad2023unsupervised, wang2025self}, but most of these methods suffer from heavy computational burdens. Obviously, this phenomenon will become more severe in 3D Transformers due to their quadratic computational complexity. To the best of our knowledge, no prior effort has attempted to apply Transformers to 3D seismic data interpolation at this stage, as training a 3D Transformer network on 3D seismic datasets demands prohibitively large computational resources.

A recent study \citep{pan2022st} demonstrates a promising approach to address the above challenges. \cite{pan2022st} proposed an image-to-video transfer learning (TL) strategy that can transfer an 2D image model to a 3D video model using parameter-efficient fine-tuning operations. Inspired by this, we propose a lightweight 2.5-dimensional (2.5D) Transformer (T-2.5D) network to reconstruct 3D incomplete seismic data without using a large amount of 3D volumes to optimize a 3D Transformer network. The proposed T-2.5D is a hybrid of 2D and 3D modules, including four 2D Transformer encoders and four 3D seismic dimension adapters (SDAs). We design a 2D-to-3D cross-dimensional TL workflow to optimize the T-2.5D. This workflow contains two stages: a 2D pre-training stage and a 3D fine-tuning stage. In the 2D pre-training stage, we use 2D training patches to just optimize the four 2D Transformer encoders without SDAs. In the 3D fine-tuning stage, we freeze the trainable parameters of the four 2D Transformer encoders and use limited 3D volumes to train the four 3D SDAs. This fine-tuning stage enables the T-2.5D to learn spatial correlation information of 3D seismic data. Overall, we use the ‘2D pre-training + 3D fine-tuning’ to replace the time-consuming and memory-intensive full 3D training, so as to significantly alleviate the computational cost of 3D Transformer-based interpolation. We investigate the interpolation performance of T-2.5D on several 3D seismic volumes. Experimental results show that the T-2.5D achieves comparable interpolation performance to 3D Transformer at a significantly lower computational cost. In other words, the proposed T-2.5D achieves a better trade-off between interpolation performance and computational cost.

\section{Methodologies}\label{sec2}
This section introduces the Transformer, Transformers with different dimensions, cross-dimensional TL, and Transformer-based interpolation theory.

\subsection{Transformer}\label{sec2.1}
As a classical DL framework, Transformer has gained significant attention from academia and industry in recent years. It has been applied to the fields of natural language processing \citep{vaswani2017attention} and computer vision \citep{dosovitskiy2020image}. Owing to the core operation, multi-head self-attention (MSA), Transformer is capable of capturing the global contextual information and shows better performance than CNNs in numerous cases.

\begin{figure}
\centering
\includegraphics[width=1\textwidth]{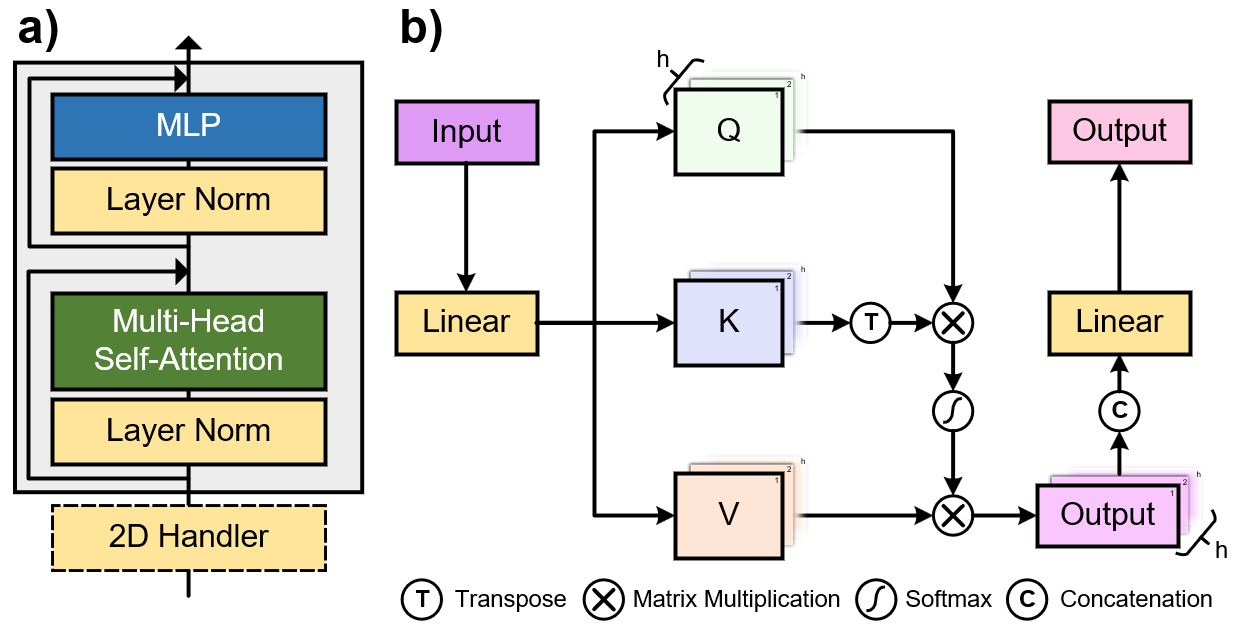}
\caption{Architecture of Transformer. (a) Transformer encoder for 2D features, and (b) calculation process of MSA.}\label{fig1}
\end{figure}

To prepare a 1D input sequence for the Transformer encoder shown in Fig. \ref{fig1}a, we first flatten the 2D input matrix and add a learnable absolute positional encoding (i.e., the 2D Handler in Fig. \ref{fig1}a). Then, the flattened feature is input into a layer normalization (LN) layer and an MSA layer, and it is also added to the original input using a residual connection. The output feature of MSA is then fed into another LN layer and a multi-layer perceptron (MLP) layer, and is added to the final output feature via the second residual connection. The output of Transformer encoders is expressed as follows:
\begin{equation}\label{eq1}
   \mitbf{X}_1=H2D(\mitbf{X})
\end{equation}
\begin{equation}\label{eq2}
   \mitbf{X}_2=MSA(LN(\mitbf{X}_1))+\mitbf{X}_1
\end{equation}
\begin{equation}\label{eq3}
   \mitbf{Y}=MLP(LN(\mitbf{X}_2))+\mitbf{X}_2
\end{equation}
where $\mitbf{X}$ and $\mitbf{Y}$ are the input and output features of Transformer encoders, and $H2D$ represents the 2D handler including the positional encoding and shape transformation. For brevity, $H2D$ or 3D handler is omitted in subsequent descriptions.

The implementation of MSA is illustrated in Fig. \ref{fig1}b. To compute the self-attention of the $i$-th head, the input feature is linearly projected into queries ($\mitbf{Q}_i$), keys ($\mitbf{K}_i$) of channel number $d_k$, and values ($\mitbf{V}_i$) of channel number $d_v$. First, we transpose the vector $\mitbf{K}_i$. Second, after the matrix multiplication between $\mitbf{Q}_i$ and $\mitbf{K}_i^T$, the product is divided by $\sqrt{d_k}$ and then input into a softmax function to generate the weights for $\mitbf{V}_i$. Finally, the attention value is obtained by multiplying $\mitbf{V}_i$ with its weights. The process is described by the following equation:
\begin{equation}\label{eq4}
   Attention_i(\mitbf{Q}_i,\mitbf{K}_i,\mitbf{V}_i)
   =softmax(\frac{\mitbf{Q}_i\mitbf{K}_i^T}{\sqrt{d_k}})\mitbf{V}_i
   .
\end{equation}
In order to fuse the information from various representation sub-spaces, the single-head attention is expanded to multi-head attention by projecting $\mitbf{Q}$s, $\mitbf{K}$s and $\mitbf{V}$s for $h$ times using separate learnable linear layers. These $h$ attentions are then integrated using concatenation operation. Finally, a linear projection is utilized to project the channel number back to the same as the input. The MSA is calculated as follows:
\begin{equation}\label{eq5}
    MSA(\mitbf{Q},\mitbf{K},\mitbf{V})
    =Linear(Concat(A_1,A_2,\ldots,A_h)),
\end{equation}
where $A_i$ is the abbreviation of $Attention_i$.

\subsection{Transformers with different dimensions}\label{sec2.2}
In this paper, we have designed three Transformer-based networks with different dimensions, including a full 2D Transformer (T-2D), a full 3D Transformer (T-3D) and a 2.5D Transformer (T-2.5D). They are highly similar in terms of architecture to ensure clear and relatively fair comparisons. Detailed descriptions of these three networks are provided below.

\subsubsection{T-2D/3D}\label{sec2.2.1}
\begin{figure*}
\centering
\includegraphics[width=1\textwidth]{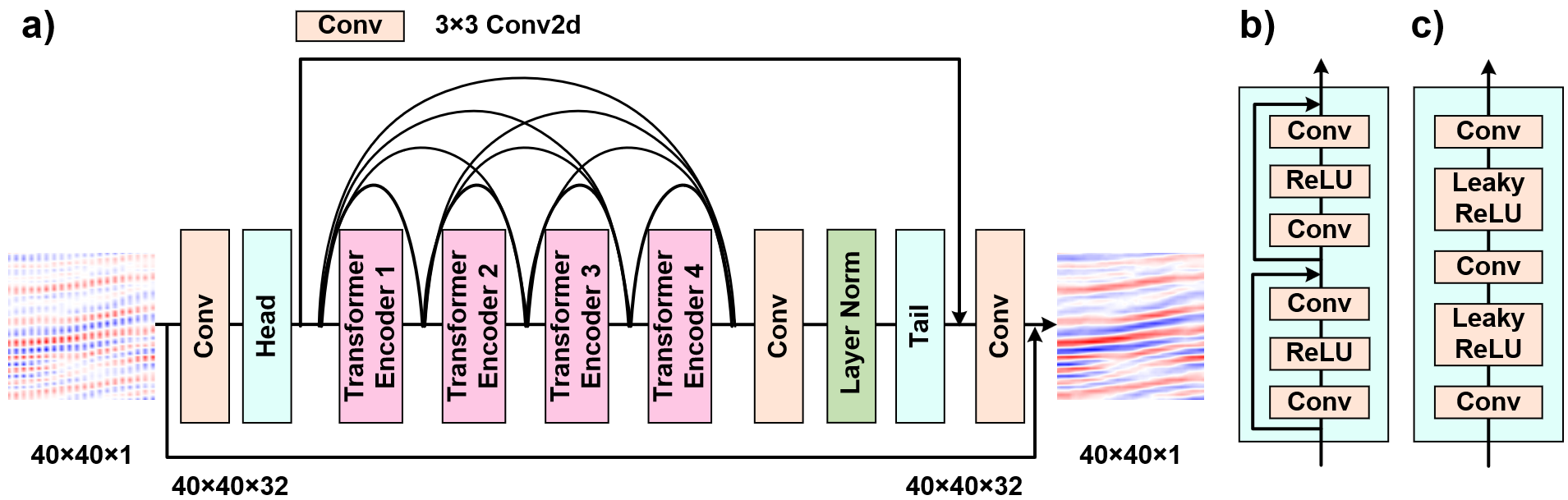}
\caption{Illustration of the T-2D. (a-c) Architectures of T-2D, HB and TB, respectively.}\label{fig2}
\end{figure*}

In Fig. \ref{fig2}a, the T-2D is composed of three 3×3 convolutional layers (Convs), a head block (HB), four 2D Transformer encoders, and a tail block (TB). A 2D patch with a size of 40×40 is firstly input into the first Conv, and the channel number is increased from 1 to 32. Then, the features with a size of 40×40×32 are input into HB whose structure is shown in Figure 2b. The HB consists of two identical parts, and each one comprises two Convs, a rectified linear unit (ReLU), and a residual connection. The output features of HB are propagated into the core module of T-2D: four successively-connected Transformer encoders, thereby capturing long-range dependencies. Moreover, dense connections \citep{Huang2017CVPR}, whose effectiveness has been validated by \cite{dong2025deep}, are deployed to enrich the feature interactions among the four Transformer encoders. Then, a Conv is used to refine the global feature, and an LN layer is deployed to stabilize the training. Subsequently, the TB displayed in Fig. \ref{fig2}c is used for final feature refinement, which is composed of three Convs interleaved with leaky ReLUs in a feed-forward manner. The TB enables efficient non-linear transformation while mitigating the risk of neuron inactivation. The output features of TB and HB are fused via a residual connection, thus avoiding the phenomenon of gradient vanishing. Finally, to generate the final output feature, the last Conv is used to integrate features across channel dimension and reduce the channel number back to one.

\begin{figure*}
\centering
\includegraphics[width=1\textwidth]{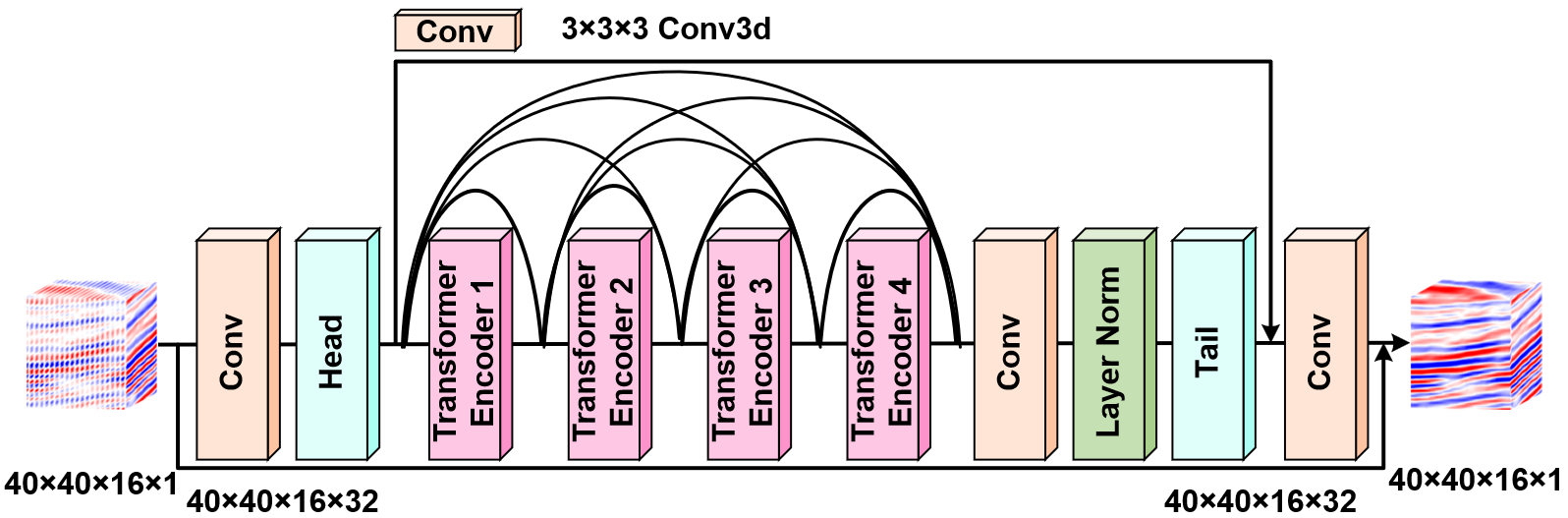}
\caption{Architecture of the T-3D, showing the changes in the dimensions of feature maps from 40×40 to 40×40×16 and network components from 2D to 3D.}\label{fig3}
\end{figure*}

We replace all replaceable layers of T-2D with their 3D counterparts to generate the T-3D shown in Fig. \ref{fig3}.

\subsubsection{T-2.5D}\label{sec2.2.2}
\begin{figure*}
\centering
\includegraphics[width=1\textwidth]{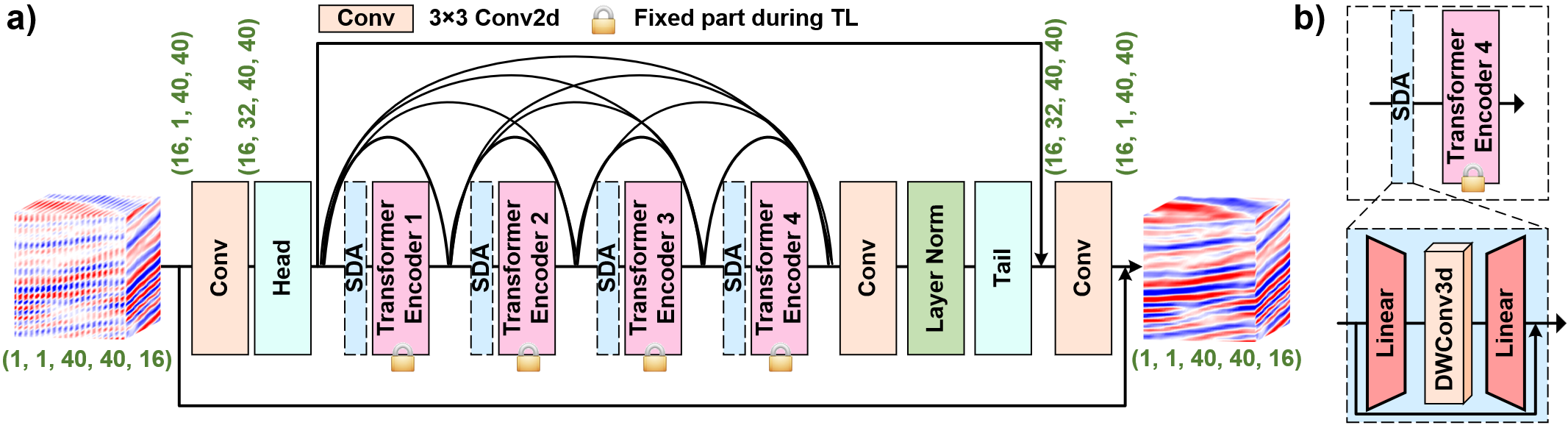}
\caption{(a) and (b) Architectures of the T-2.5D and SDA, respectively.}\label{fig4}
\end{figure*}

To keep a low computational cost, we kept most of the 2D components unchanged, and design T-2.5D by simply adding four 3D SDAs to T-2D. In our implementation, a 3D volume is represented as a tensor of shape (B, C, T, X, I), where the five dimensions denote the batch size, channel number, time sample, crossline point, and inline point, respectively. As shown in Fig \ref{fig4}a, at the beginning of T-2.5D, the tensor is permuted to a shape of (B, I, C, T, X) by reordering its dimensions. Subsequently, the B and I dimensions are merged, resulting in a tensor of shape (B×I, C, T, X). This reshaping makes the tensor compatible with the input requirements of the 2D network components. The tensor shapes are labeled in dark green in Fig. \ref{fig4}a. In Fig. \ref{fig4}a, we place a 3D SDA before each 2D Transformer encoder. As shown in Fig. \ref{fig4}b, the 3D SDA is composed of two linear layers and a depth-wise 3D Conv (DWConv3d). Specifically, the first linear layer is used to project the input feature into a 3D space, allowing the following DWConv3d to capture abundant spatial contextual information. The tensor is restored to the shape (B, C, T, X, I) only before being input to DWConv3d, and is converted back to (B×I, C, T, X) immediately after DWConv3d. Another linear layer is used to project the feature back to the original dimension. A residual connection is utilized to send the input to the output, to prevent potential information loss.

\subsection{Cross-dimensional TL}\label{sec2.3}
\begin{figure}
\centering
\includegraphics[width=1\textwidth]{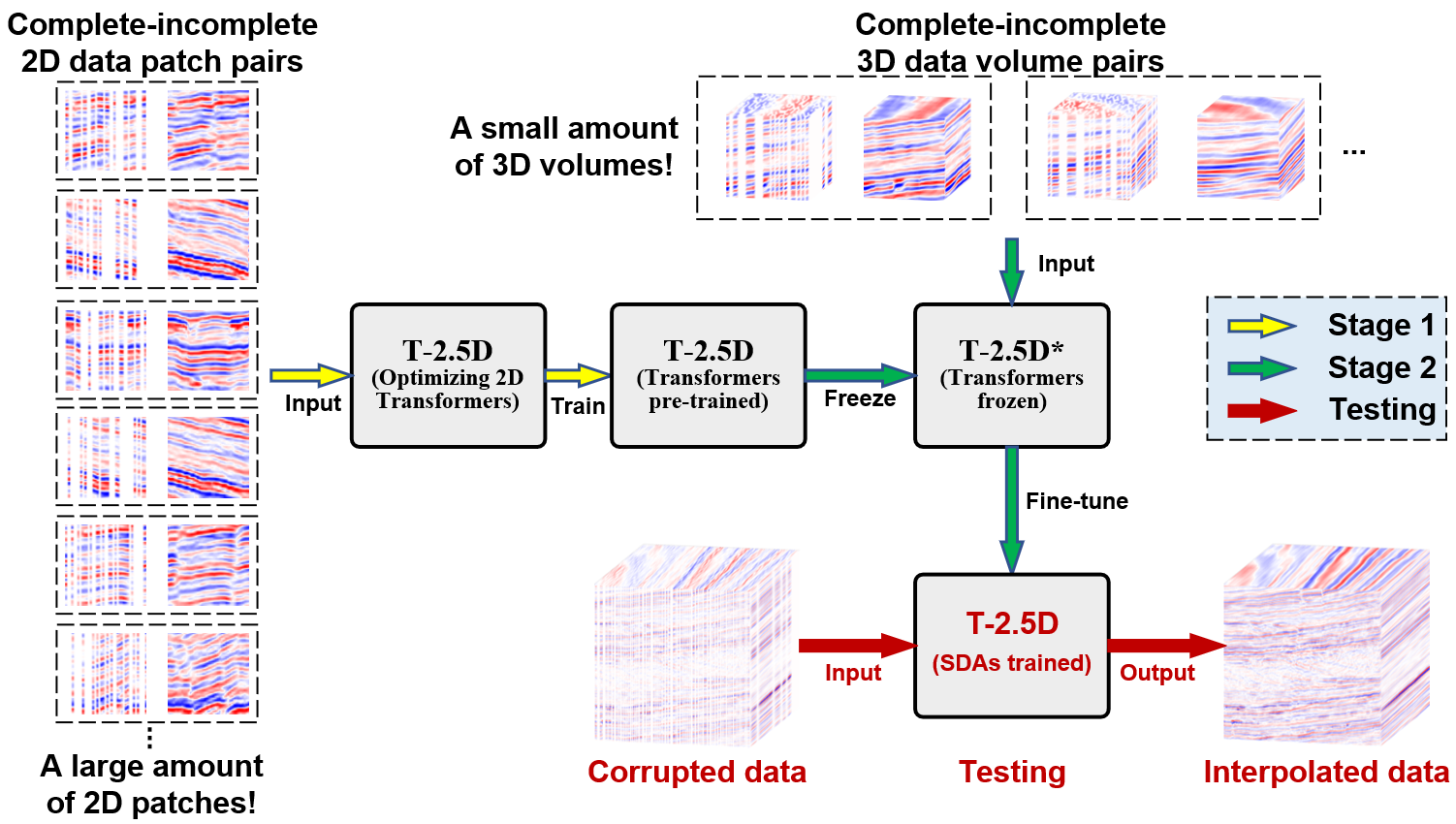}
\caption{Cross-dimensional TL of the T-2.5D.}\label{fig5}
\end{figure}

We utilize a cross-dimensional TL strategy to optimize the T-2.5D. This workflow comprises two stages, including a 2D pre-training one and a 3D fine-tuning one. In the first stage, we optimize the four 2D Transformer encoders using a large number of 2D patches, so as to generate a 2D pre-trained model. In the second stage, we freeze the trainable parameters of the four Transformer encoders and use a small amount of 3D volumes to optimize the four 3D SDAs of T-2.5D, where spatial information across seismic lines is learned.The overall process of the cross-dimensional TL is provided in Fig. \ref{fig5}.

\subsection{Transformer-based Interpolation Theory}\label{sec2.4}
\subsubsection{2D/3D Interpolation}\label{sec2.4.1}
For the complete 2D seismic data $\mitbf{x}$, its corresponding incomplete 2D data $\mitbf{y}$ can be expressed as:
\begin{equation}\label{eq6}
   \mitbf{y}=\mitbf{Ax}
   ,
\end{equation}
where $\mitbf{A}$ represents a 2D masking matrix used to generate the decimated data with missing traces. Through the network training of T-2D, we can establish a nonlinear mapping relationship $T_{2D}$ between $\mitbf{x}$ and $\mitbf{y}$. The trainable parameters of T-2D are represented as $\theta_{2D}=\{\mitbf{\omega},\mitbf{b}\}$, where $\mitbf{\omega}$ and $\mitbf{b}$ denote the weights and biases, respectively. $\theta_{2D}$ is updated by minimizing the L\textsubscript{2}-norm loss function in Eq. (\ref{eq7}).
\begin{equation}\label{eq7}
   l(\theta_{2D})=\frac{1}{B}\sum_{i=1}^{B}{\lVert{T_{2D}(\mitbf{y}_i,\theta_{2D})-\mitbf{x}_i}\rVert}^2
   ,
\end{equation}
where $B$ is the batch size (i.e., the pair number of training data patches used in one iteration), $\mitbf{x}_i$ and $\mitbf{y}_i$ represent the $i$-th paired training data patches, and $\lVert\cdot\rVert$ denotes the L\textsubscript{2} norm. In this paper, we use the AdamW algorithm \citep{loshchilov2017decoupled} to update the trainable parameters in an iterative manner. Upon obtaining the optimal parameters $\theta_{2D}^{opt}$, we can generate the interpolated result  ${\hat{\mitbf{x}}}_{opt}$ by inputting the incomplete data $\mitbf{y}$ into the nonlinear relationship $T_{2D}$:
\begin{equation}\label{eq8}
   {\hat{\mitbf{x}}}_{opt}=T_{2D}(\mitbf{y},\theta_{2D}^{opt})
   .
\end{equation}

The principle of 3D interpolation is similar to the 2D case, where we utilize 3D data to optimize the 3D trainable parameters of T-3D.

\subsubsection{2.5D Interpolation}\label{sec2.4.2}
In this paper, we utilize two-stage training strategy to optimize the trainable parameters of T-2.5D, $\theta_{2.5D}$, so as to achieve the 2.5D interpolation using Transformer. The $\theta_{2.5D}$ is composed of $\theta_{2D}^\prime$, $\theta_{3D}^\prime$, representing the trainable parameters of the 2D Transformer encoders and 3D SDAs, respectively.

In the 2D pre-training stage, the loss function is updated as follows:
\begin{equation}\label{eq9}
   l_{stage1}(\theta_{2D}^\prime)=\frac{1}{B}
   \sum_{i=1}^{B}{\lVert{T_{2.5D}(\mitbf{y}_i^{2D},\theta_{2D}^\prime)-\mitbf{x}_i^{2D}}\rVert}^2
   ,
\end{equation}
where $\mitbf{x}_i^{2D}$ and $\mitbf{y}_i^{2D}$ denote the $i$-th paired 2D training data patches, respectively. Upon obtaining the optimal parameters $\theta_{2D}^{\prime opt}$, we freeze them and use a small amount of 3D volumes to optimize the 3D SDAs. In the 3D fine-tuning stage, the loss function is updated through Eq. (\ref{eq10}).
\begin{equation}\label{eq10}
   l_{stage2}(\theta_{2.5D})=\frac{1}{B}
   \sum_{i=1}^{B}{\lVert{T_{2.5D}(\mitbf{y}_i^{3D},\{\bar{\theta}_{2D}^\prime,\theta_{3D}^\prime\})-\mitbf{x}_i^{3D}}\rVert}^2
    ,
\end{equation}
where $\mitbf{x}_i^{3D}$ and $\mitbf{y}_i^{3D}$ represent the i-th paired 3D training data volumes, respectively, and ${\bar{\theta}}_{2D}^\prime$ denotes the frozen parameters of 2D Transformer encoders. After certain iterations of the fine-tuning stage, we can obtain a set of optimal 2.5D parameters $\theta_{2.5D}^{opt}$. The final 3D interpolation result  ${\hat{\mitbf{x}}}_{opt}^{3D}$ is generated by inputting the incomplete 3D data $\mitbf{y}^{3D}$ into the 2.5D nonlinear relationship $T_{2.5D}$ as follows:
\begin{equation}\label{eq11}
   {\hat{\mitbf{x}}}_{opt}^{3D}=T_{2.5D}(\mitbf{y}^{3D},\theta_{2.5D}^{opt})
   .
\end{equation}
The specific process of 2.5D interpolation is given in Algorithm \ref{alg1}.
\begin{algorithm}
\small
\caption{2.5D Interpolation}\label{alg1}
\begin{algorithmic}[1]
\REQUIRE $T_{2.5D}$, the nonlinear relationships of 2.5D Transformer; $B$, batch size; $E_1$ and $E_2$, the numbers of epochs for the two stages, respectively; $K$, the number of iterations in each epoch; $D_{2D}$ and $D_{3D}$, 2D and 3D complete datasets, respectively; $\mitbf{y}$, the 3D volume to be tested.
\STATE \textbf{STAGE 1: 2D PRE-TRAINING.}
\FOR {i=1,2,...,$E_1$}
    \FOR {j=1,2,...,$K$}
        \STATE Sample $C_{2D}=\{\mitbf{x_k}|k=1,2,\ldots,B\}$, a batch of complete data patches, from $D_{2D}$.
        \STATE Sample ${Ic}_{2D}=\{\mitbf{y_k^n}|\mitbf{y_k^n}=RM(\mitbf{x_k^n}),\mitbf{x_k^n}\in C_{2D},k=1,2,\ldots,B\}$, a batch of incomplete data patches from $C_{2D}$, where $RM$ represents the operation of removing traces.
        \STATE Loss iteration: $\theta_{2D}^\prime\gets{\nabla}_\theta[\frac{1}{B}\sum^{B}_{k=1}{\lVert T_{2D}(\mitbf{y_k^n},\theta_{2D}^\prime)-\mitbf{x_k^n}\rVert}^2]
        ,\mitbf{x_k^n}\in C_{2D}^n, \mitbf{y_k^n}\in {Ic}_{2D}$.
    \ENDFOR
\ENDFOR
\STATE Optimal parameters obtained: $\theta_{2D}^{\prime opt}$.
\STATE \textbf{STAGE 2: 3D FINE-TUNING.}
\STATE 	Get ${\bar{\theta}}_{2D}^\prime$, the frozen $\theta_{2D}^\prime$.
\FOR {i=1,2,...,$E_2$}
    \FOR {j=1,2,...,$K$}
    \STATE Sample $C_{3D}=\{\mitbf{x_k}|k=1,2,\ldots,B\}$, a batch of complete data patches, from $D_{3D}$.
    \STATE Sample ${Ic}_{3D}=\{\mitbf{y_k^n}|\mitbf{y_k^n}=RM(\mitbf{x_k^n}),\mitbf{x_k^n}\in C_{3D},k=1,2,\ldots,B\}$, a batch of incomplete data volumes from $C_{3D}$, where $RM$ represents the operation of removing traces.
    \STATE Loss iteration: $\theta_{2.5D}\gets{\nabla}_\theta[\frac{1}{B}\sum^{B}_{k=1}{\lVert T_{2.5D}(\mitbf{y_k^n},\{\bar{\theta}_{2D}^\prime,\theta_{3D}^\prime\})-\mitbf{x_k^n}\rVert}^2]
        ,\mitbf{x_k^n}\in C_{3D}^n, \mitbf{y_k^n}\in {Ic}_{3D}$, where $\theta_{3D}^\prime$ is the trainable parameters of the four SDAs.
    \ENDFOR
\ENDFOR
\STATE Optimal parameters obtained: $\theta_{2.5D}^{opt}$.
\STATE \textbf{INTERPOLATION.}
\STATE ${\hat{\mitbf{x}}}_{opt}^{3D}\gets T_{2.5D}(\mitbf{y}^{3D},\theta_{2.5D}^{opt})$, ${\hat{\mitbf{x}}}_{opt}^{3D}$ is the final interpolated result.
\end{algorithmic}
\end{algorithm}

\section{Experiments}\label{sec3}
\subsection{Training and Hyperparameters Settings}\label{sec3.1}
The training program is executed on Pytorch 2.10.0 and CUDA 12.8, running on an Ubuntu 22.04 operation system. Hardware configurations consist of an NVIDIA A800 GPU with 80GB of memory, an Intel(R) Xeon(R) Gold 6348 CPU with 28 cores at 2.6GHz frequency, and 120GB RAM.

Taking into account performance, computational resources, and fair comparison, we set the hyperparameters as shown in Table \ref{tab1}. The models are trained using the AdamW optimizer with an initial learning rate setting of $1\times10^{-4}$, a weight decay of $1\times10^{-5}$, and default momentum coefficients of 0.9 and 0.999. A cosine learning rate scheduler with linear warmup is employed, with the learning rate updated after each training iteration. Stage 1 consists of 20 epochs with 6,000 iterations per epoch (120,000 iterations in total), while Stage 2 consists of 20 epochs with 2,000 iterations per epoch (40,000 iterations in total). In Stage 1, the learning rate increases linearly from 0 to $1\times10^{-4}$ during the first 6,000 iterations and subsequently decreases to 0 following a cosine schedule. Similarly, in Stage 2, the learning rate increases linearly from 0 to $1\times10^{-4}$ during the first 2,000 iterations and subsequently decreases to 0 following a cosine schedule. A dropout rate of 0.1 is applied at two locations: immediately before the final matrix multiplication in the MSA calculation and at the end of the MLP. The checkpoint achieving the highest product of PSNR and SSIM on the validation set is selected as the optimal model.

Regarding data preprocessing, each seismic volume is normalized by dividing its amplitude values by the maximum absolute value before patch/volume extraction. Then, we perform random masking of 40\%–60\% of the traces in each patch/volume to create complete–incomplete data pairs. In the binary mask, 1 indicates retained data, while 0 indicates missing data. The mask is applied trace-wise, meaning that all samples within a trace share the same mask value (either 0 or 1), with no mixture of the two values within a single trace. During testing, the test data are similarly normalized using their maximum absolute value before being fed into the model. After inference, inverse normalization is performed by multiplying the interpolated data by the corresponding normalization factor to restore the original amplitude scale.
\begin{table}
\caption{Hyperparameters of T-2D, T-3D, and T-2.5D.}\label{tab1}%
\begin{tabular}{@{}lc@{}}
\toprule
Hyperparameters & Specifications  \\
\midrule
Optimizer               & AdamW    \\
Loss function           & L2 norm with mean reduction    \\
Data patch size         & 40×40 (T-2D), 40×40×16 (T-3D and T-2.5D)    \\
Batch size              & 1         \\
Number of epochs        & 20         \\
Initial Learning rate           & 10\textsuperscript{-4}  \\
Scheduler               & Cosine scheduler with warmup  \\
Input channel number    & 1         \\
Embedding channel number& 32         \\
Total layers            & 4         \\
\bottomrule
\end{tabular}
\end{table}

\subsection{Metrics for Interpolation Results}\label{sec3.2}
In this study, the peak signal-to-noise ratio (PSNR) and structural similarity index measure \citep[SSIM;][]{wang2004image} are used to evaluate the performance of different methods quantitatively. The PSNR of 2D data is defined as the following equation:
\begin{equation}\label{eq12}
   PSNR=10\log_{10}\left\{\frac{[max(abs(\boldsymbol{x}))]^2}
   {\frac{1}{MN}\sum_{i=0}^{M-1}\sum_{j=0}^{N-1}
   [\boldsymbol{x}(i,j)-\boldsymbol{y}(i,j)]^2}\right\}
   ,
\end{equation}
where $\boldsymbol{x}$ and $\boldsymbol{y}$ denote the complete and interpolated data, respectively, $M$ represents the number of sample points in each trace, $N$ is number of traces, and $max(abs(\boldsymbol{x}))$ represents the maximum absolute value of $\boldsymbol{x}$. Similarly, the 3D version of PSNR is expressed as follows:
\begin{equation}\label{eq13}
   PSNR=10\log_{10}\left\{\frac{[max(abs(\boldsymbol{x}))]^2}{
   \frac{1}{SXI}\sum_{i=0}^{S-1}\sum_{j=0}^{X-1}\sum_{k=0}^{I-1}
   [\boldsymbol{x}(i,j,k)-\boldsymbol{y}(i,j,k)]^2}\right\}
   ,
\end{equation}
where $S$, $X$, and $I$ denote the number of sampled points along the time, crossline, and inline axes, respectively.

SSIM is a commonly used metric to evaluate the perceptual similarity between two data. It considers the changes in structural information, luminance, and contrast \citep{wang2004image}. Given a 2D/3D complete data $\boldsymbol{x}$ and its corresponding interpolated result $\boldsymbol{y}$, SSIM is defined as:
\begin{equation}\label{eq14}
   SSIM(\boldsymbol{x},\boldsymbol{y})=\frac{
   (2\mu_{\boldsymbol{x}}\mu_{\boldsymbol{y}}+C_1)(2\sigma_{\boldsymbol{xy}}+C_2)}
   {(\mu_{\boldsymbol{x}}^2+\mu_{\boldsymbol{y}}^2+C_1)(\sigma_{\boldsymbol{x}}^2+\sigma_{\boldsymbol{y}}^2+C_2)}
   ,
\end{equation}
where $\mu_{\boldsymbol{x}}$ and $\mu_{\boldsymbol{y}}$ represent the means of $\boldsymbol{x}$ and $\boldsymbol{y}$, respectively, $\sigma_{\boldsymbol{x}}^2$ and $\sigma_{\boldsymbol{y}}^2$ are the variances of $\boldsymbol{x}$ and $\boldsymbol{y}$, respectively, $\sigma_{\boldsymbol{xy}}$ is the covariance between $\boldsymbol{x}$ and $\boldsymbol{y}$, and $C_1$, $C_2$ are small constants to stabilize the division of whole equation. The value of SSIM ranges from 0 to 1, with higher value indicating greater similarity.

The loss, PSNR and SSIM values are calculated over the entire reconstructed patch/volume against the corresponding complete label data.

\subsection{Test of Kerry Dataset}\label{sec3.3}
\subsubsection{Data preparation and training}\label{sec3.3.1}
\begin{figure}
\centering
\includegraphics[width=0.5\textwidth]{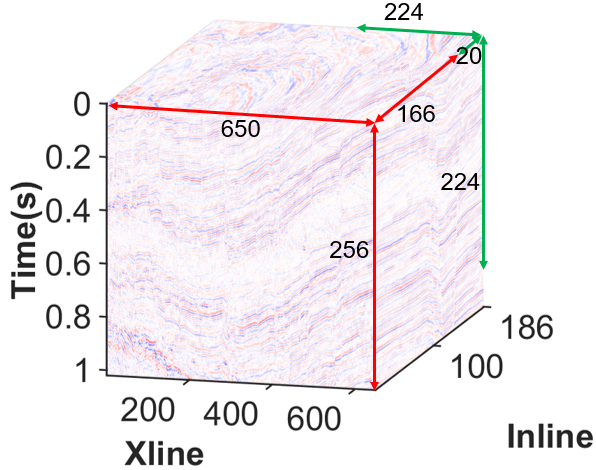}
\caption{
3D Volume of the Kerry dataset.
}\label{fig6}
\end{figure}

We first use the Kerry dataset, which is a 3D marine survey offshore New Zealand \citep{Kerry3D}. The original Kerry dataset contains one 3D volume of 1252 time samples, 735 crossline points, and 287 inline points. The time sampling interval is 0.004 s. We extract from it a 3D volume with a shape of 256×650×186, as shown in Fig. \ref{fig6}. The training and test blocks are defined by red and green double arrows in Fig. \ref{fig6}, respectively. The training block is only used to generate training and validation sets, and the test block is only used for testing. To generate sufficient data patches/volumes, we adopt a sliding window technique to extract a 2D and a 3D collections of patches/volumes from the training block. Then, the training and validation sets are randomly sampled from the above collections. The Kerry dataset, along with the Parihaka, Opunake, and Stratton datasets and the Opunake velocity model used later in this study, are all standard publicly available datasets from SEG. Further details can be found in the corresponding references.

\begin{figure}
\centering
\includegraphics[width=1\textwidth]{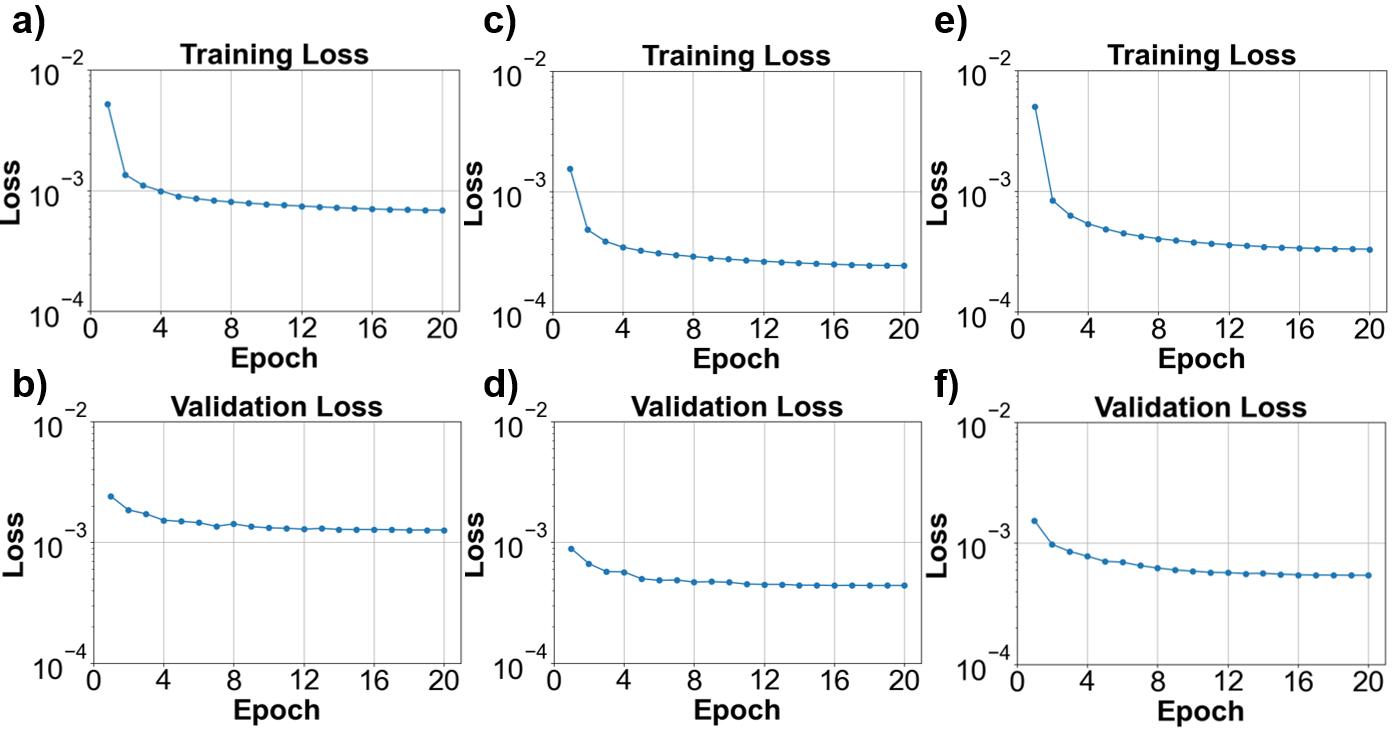}

\caption{
L2 loss curves on Kerry dataset. (a, c, and e) Training loss curves of T-2D, T-3D and T-2.5D, respectively; and (b, d, and f) the corresponding validation loss curves.
}\label{fig7}
\end{figure}

For T-2D, we sample 6000 and 1500 patches, with a shape of 40×40, from the 2D collection as training and validation sets. Similarly, for T-3D, 6000 and 1500 volumes (40×40×16) are sampled from the 3D collection as training and validation sets. In the first stage of T-2.5D, we pre-train the 2D Transformer encoders using the same dataset as in T-2D. The second stage of T-2.5D utilizes 2000 and 500 volumes sampled from the 3D collection as training and validation sets, respectively, to fine-tune the 3D SDAs. In each patch/volume, 40\%–60\% of the traces are randomly removed to generate complete-incomplete data pairs for supervised training. After training, the L2 loss curves for both training and validation of the three methods are provided in Fig. \ref{fig7}a–\ref{fig7}f.

During the testing process, a sliding window technique is also adopted to ensure that the entire 224×224×20 test block is fully covered. The sliding window is moved with a stride equal to 50\% of the patch/volume size. For T-2D, the test block is processed line by line using the sliding window with a stride of 20×20. For T-2.5D and T-3D, the stride is 20×20×8. Predictions in overlapping regions were averaged.

\subsubsection{Testing results}\label{sec3.3.2}
\begin{figure}
\centering
\includegraphics[height=0.9\textheight]{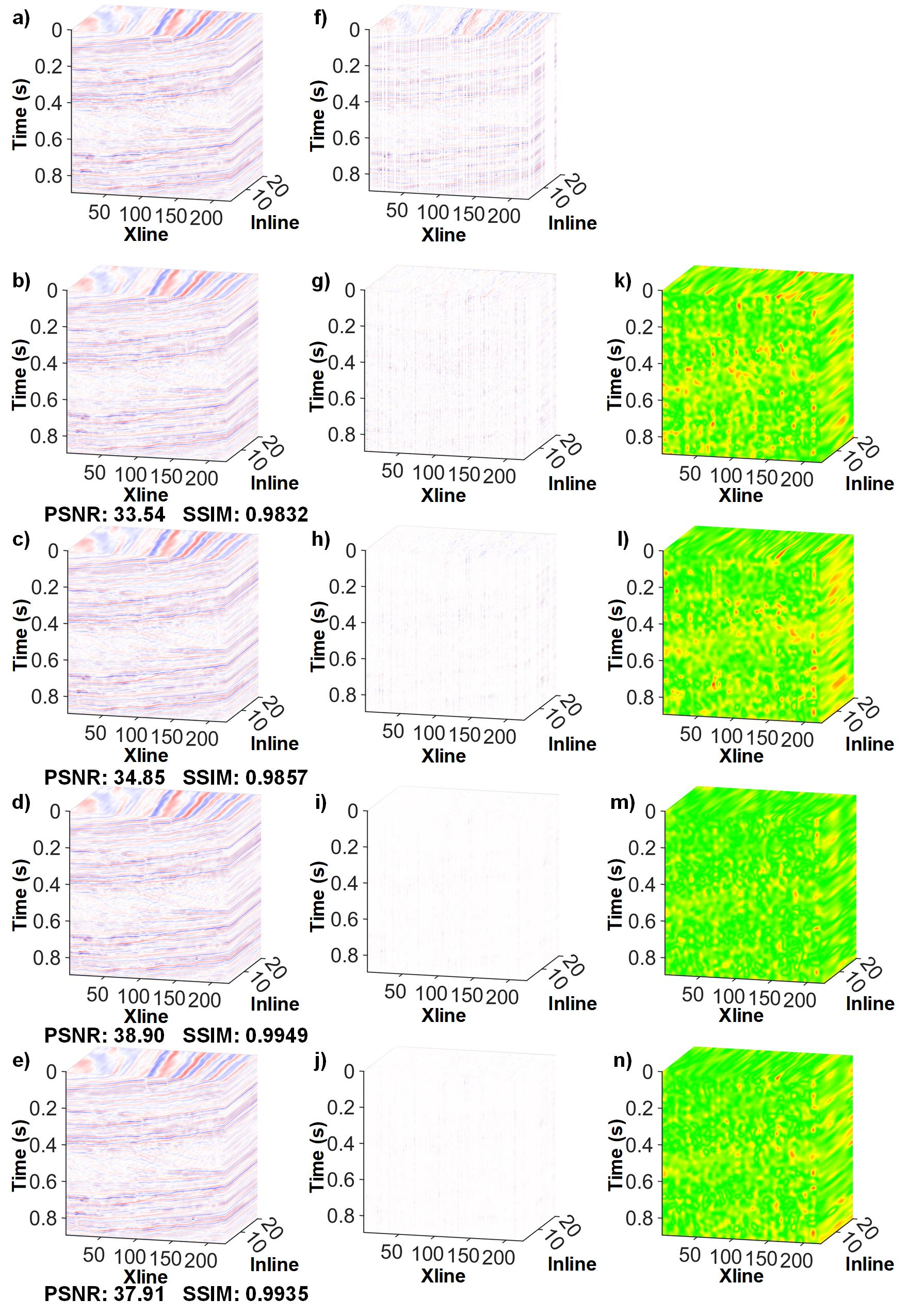}
\caption{
Interpolation results of Kerry dataset. (a) Complete data, (f) 50\% randomly sampled data, (b–e) interpolated results of POCS, T-2D, T-3D and T-2.5D, respectively; (g–j) the corresponding residual images, and (k–n) local similarity maps between (g–j) and (a).
}\label{fig8}
\end{figure}

We additionally select a classical POCS-based interpolation method \citep{abma20063d} as an additional benchmark for comparison. The initial threshold parameters is set to 0.8, and the algorithm lasts for 300 iterations. As shown in Fig. \ref{fig8}a, the test data is of shape 224×224×20 extracted from Fig. \ref{fig6}. We randomly delete 50\% of the traces to generate the incomplete data in Fig. \ref{fig8}f. Fig. \ref{fig8}b–\ref{fig8}e present the interpolated results. Fig. \ref{fig8}g–\ref{fig8}j display the residual images between the interpolated results and the complete data, and Fig. \ref{fig8}k–\ref{fig8}n are the local similarity \cite{fomel2007local} maps between Fig. \ref{fig8}g–\ref{fig8}j and Fig. \ref{fig8}a. Both of them are utilized to measure the signal leakage of the interpolated results, and lower energy indicates better interpolation performance. All the four methods are able to reconstruct the missing traces to some extent. However, the result of POCS in Fig. \ref{fig8}b exhibits discontinuous events. Stronger signal leakage can be observed in the results of POCS and T-2D in Fig. \ref{fig8}g, \ref{fig8}h, \ref{fig8}k, and \ref{fig8}l. In comparison, T-2.5D achieves a comparable performance to T-3D, which validates the effectiveness of T-2.5D.

\begin{figure}
\centering
\includegraphics[width=1\textwidth]{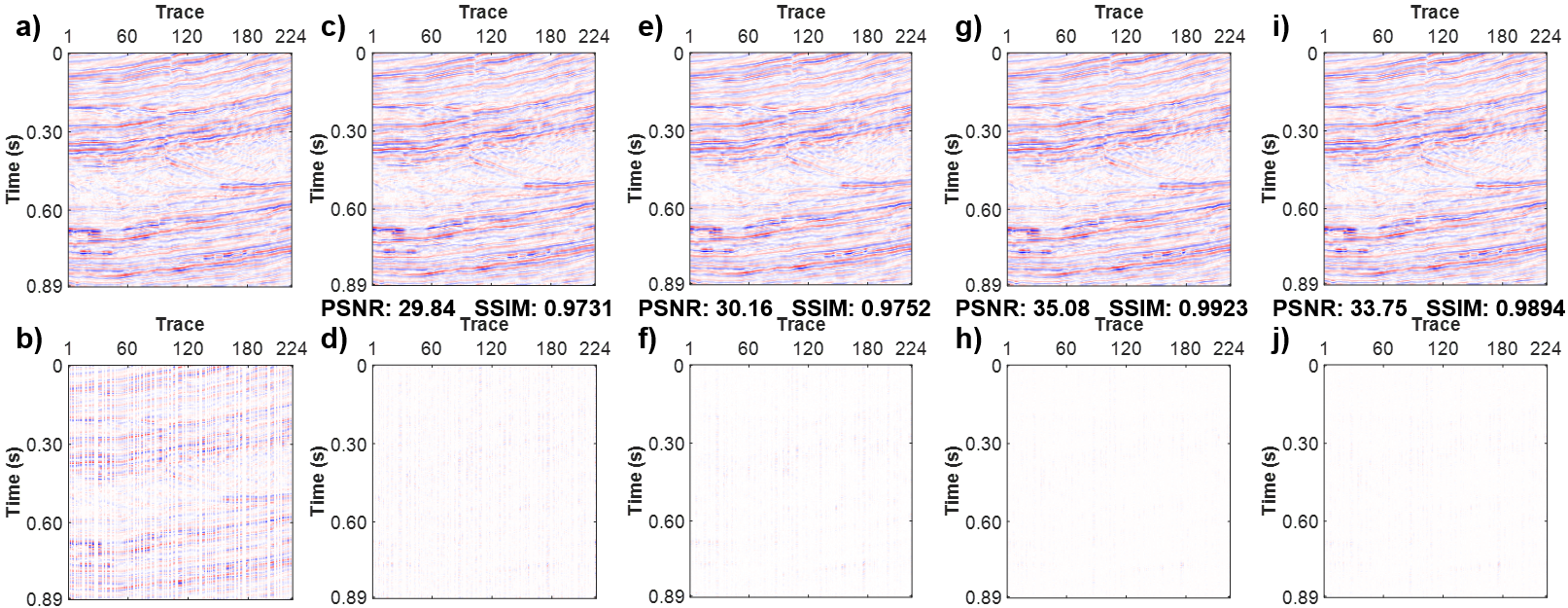}
\caption{
Comparisons of the sixth line extracted from Fig. \ref{fig8}. (a) Complete data, (b) 50\% randomly sampled data, (c, e, g, and i) interpolated results of POCS, T-2D, T-3D, and T-2.5D, respectively; and (d, f, h, and j) the corresponding residual images.
}\label{fig9}
\end{figure}

For a clearer comparison, the sixth line is plotted in 2D (Fig. \ref{fig9}). As shown in Fig. \ref{fig9}c, POCS struggle to interpolate missing traces for its lowest numerical result. Due to the lack of spatial correlation information, T-2D also yields poor performance. These are evidenced by their significantly lower quantitative results and more severe signal leakage in Fig. \ref{fig9}d and \ref{fig9}f. In comparison, T-2.5D achieves an interpolation result close to that of T-3D, and both of their residual images show visually little signal leakage.

\begin{figure}
\centering
\includegraphics[width=1\textwidth]{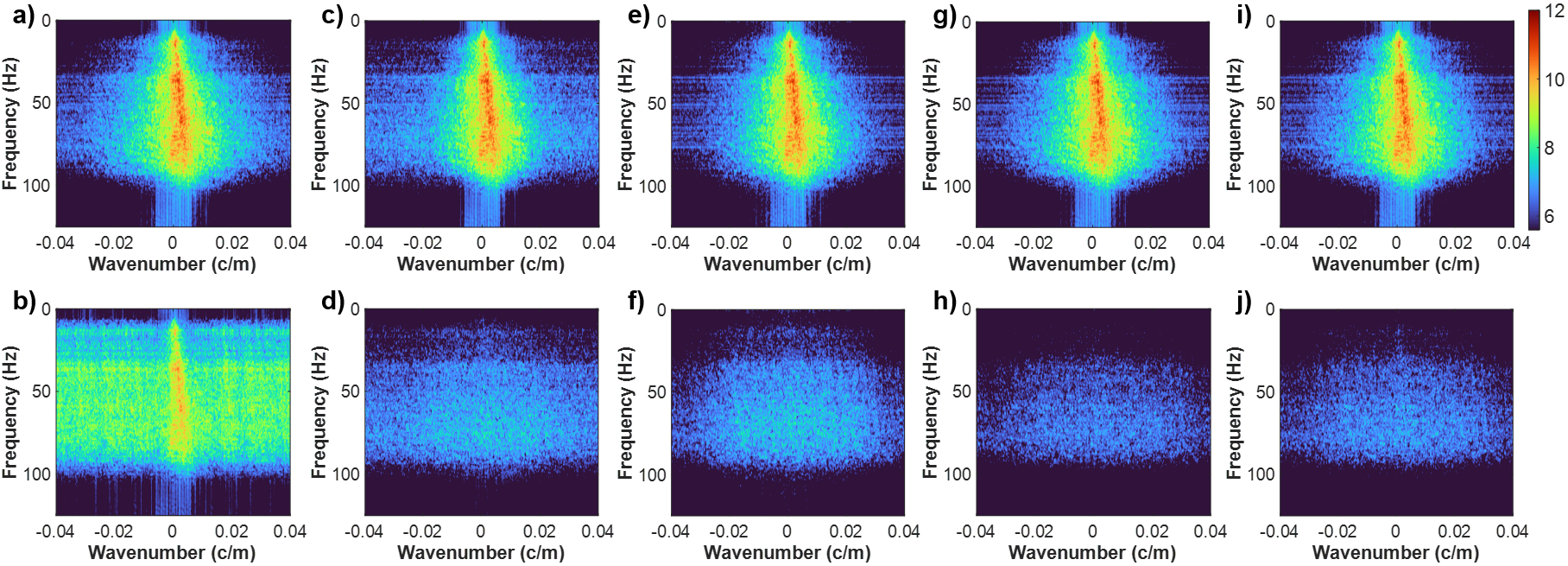}
\caption{
(a and b) f-k spectra of the complete data (Fig. \ref{fig9}a) and 50\% randomly sampled data (Fig. \ref{fig9}b), (c, e, g, and i) f-k spectra of the interpolated results by POCS, T-2D, T-3D and T2.5D (Fig. \ref{fig9}c, \ref{fig9}e, \ref{fig9}g, and \ref{fig9}i), respectively; and (d, f, h and j) f-k spectra of the corresponding residual images (Fig. \ref{fig9}d, \ref{fig9}f, \ref{fig9}h, and \ref{fig9}j).
}\label{fig10}
\end{figure}

Furthermore, we plot in Fig. \ref{fig10} the f-k spectra of the complete data (Fig. \ref{fig9}a), 50\% randomly sampled data (Fig. \ref{fig9}b), the four interpolated results (Fig. \ref{fig9}c, \ref{fig9}e, \ref{fig9}g, and \ref{fig9}i), and the corresponding residual images (Fig. \ref{fig9}d, \ref{fig9}f, \ref{fig9}h, and \ref{fig9}j). As illustrated in Fig. \ref{fig10}c, \ref{fig10}e, \ref{fig10}g, and \ref{fig10}i, the aliased energy presented in corrupted data (Fig. \ref{fig10}b) has been largely removed by all the four methods. However, as shown in Fig. \ref{fig10}d and \ref{fig10}f, stronger residual interference exists in the results of POCS and T-2D. It is also observed in Fig. \ref{fig10}j that the energy leakage of T-2.5D is evidently weaker than that of T-2D, and is very close to that of T-3D in Fig. \ref{fig10}h.

\subsubsection{Analysis of the computational cost}\label{sec3.3.3}
Computational cost plays a crucial role in DL-based methods as it directly reflects the efficiency of different methods. We have recorded the peak GPU memory usage and training time of the three networks in Table \ref{tab2}. Notably, the total training time of T-2.5D is composed of the pre-training time of the first stage and the training time of the second stage. T-2.5D requires only about 1/12 the memory of T-3D with the training time reduced to approximately 1/20. Generally, T-2.5D achieves comparable interpolation performance to T-3D, while substantially reducing the memory usage and training-time cost, demonstrating that T-2.5D is an efficient DL-based seismic data interpolation method and the proposed cross-dimensional TL is a lightweight process.
\begin{table}
\caption{Computational costs of T-2D, T-3D, and T-2.5D on the Kerry dataset.}\label{tab2}%
\begin{tabular}{@{}lcc@{}}
\toprule
Method & Peak GPU memory usage (MB) & Training time (h)  \\
\midrule
T-2D            & 776           & 0.59    \\
T-3D            & 56852         & 21.06    \\
T-2.5D & 4636 & 0.59 (Stage 1) + 0.65 (Stage 2)    \\
\bottomrule
\end{tabular}
\end{table}

\subsection{Test on Parihaka Dataset}\label{sec3.4}
\subsubsection{Data preparation and training}\label{sec3.4.1}
\begin{figure}
\centering
\includegraphics[width=0.5\textwidth]{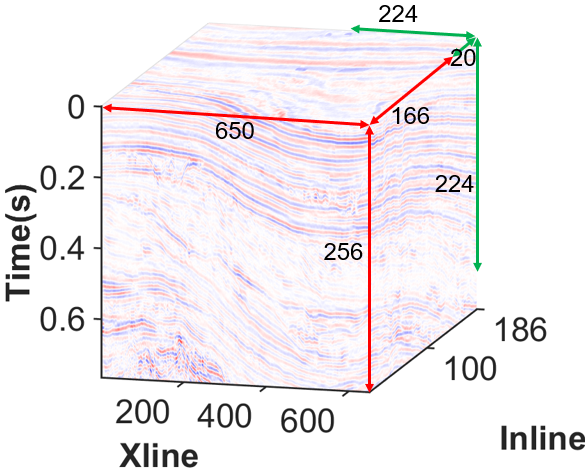}
\caption{
3D Volume of the Parihaka dataset.
}\label{fig11}
\end{figure}

\begin{figure}
\centering
\includegraphics[width=1\textwidth]{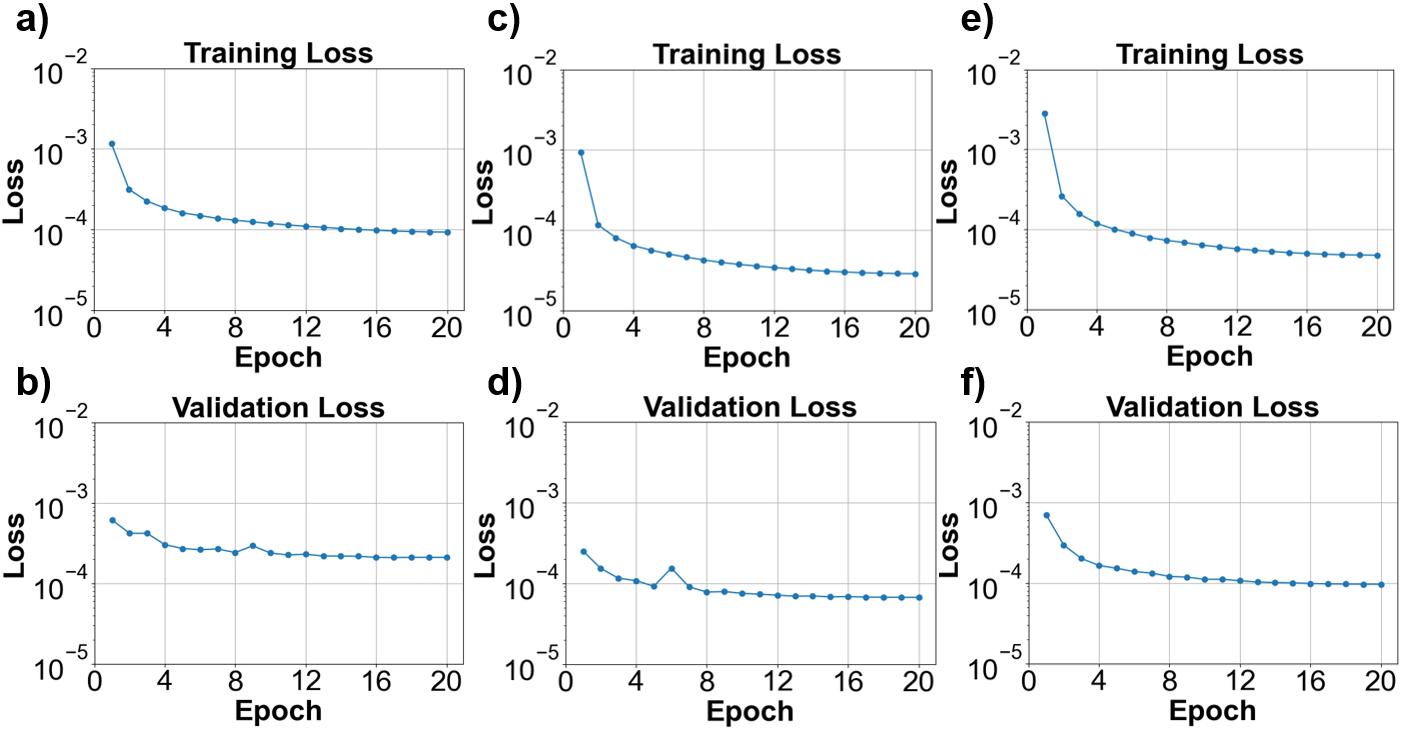}
\caption{
L2 loss curves on Parihaka dataset. (a, c, and e) Training loss curves of T-2D, T-3D and T-2.5D, respectively; and (b, d, and f) the corresponding validation loss curves.
}\label{fig12}
\end{figure}

In this subsection, we use the Parihaka dataset obtained from another survey area of New Zealand \citep{Parihaka3D} to further validate the effectiveness of T-2.5D. The time sampling interval is 0.003 s. We extract a 3D volume with a shape of 256×650×186, as shown in Fig. \ref{fig11}, and its dataset preparations and testing process are the same as those of the Kerry dataset. Fig. \ref{fig12}a–\ref{fig12}f shows the convergence behavior of the training and validation curves of the three methods on the Parihaka dataset.

\subsubsection{Testing results}\label{sec3.4.2}
\begin{figure}
\includegraphics[height=0.9\textheight]{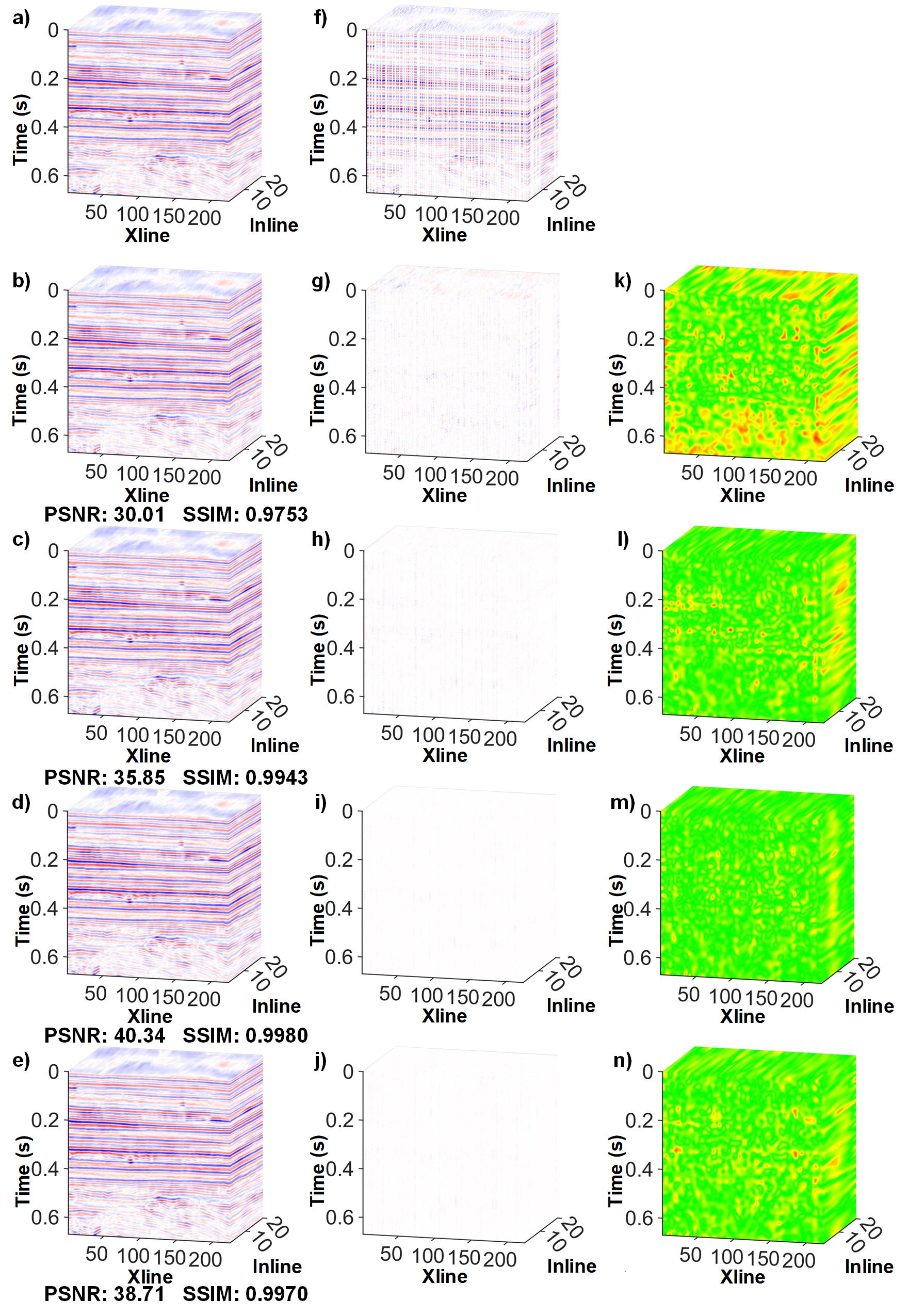}
\caption{
Interpolation results of Parihaka dataset. (a) Complete data, (f) 50\% randomly sampled data, (b–e) interpolated results of POCS, T-2D, T-3D and T-2.5D, respectively; (g–j) the corresponding residual images, and (k–n) local similarity maps between (g–j) and (a)..
}\label{fig13}
\end{figure}

The 3D data volume (224×224×20) defined by green double arrows in Fig. \ref{fig11} is likewise extracted for testing. As shown in Fig. \ref{fig13}b–\ref{fig13}e, the performances of POCS and T-2D remain markedly inferior to those of T-3D and T-2.5D, which are also supported by the strongest signal leakage in their residual images (Fig. \ref{fig13}g and \ref{fig13}h) and local similarity map (Fig. \ref{fig13}k and \ref{fig13}l). In contrast, the performance of T-2.5D is very close to that of T-3D, as shown in Fig. \ref{fig13}j and \ref{fig13}n.

\begin{figure}
\centering
\includegraphics[width=1\textwidth]{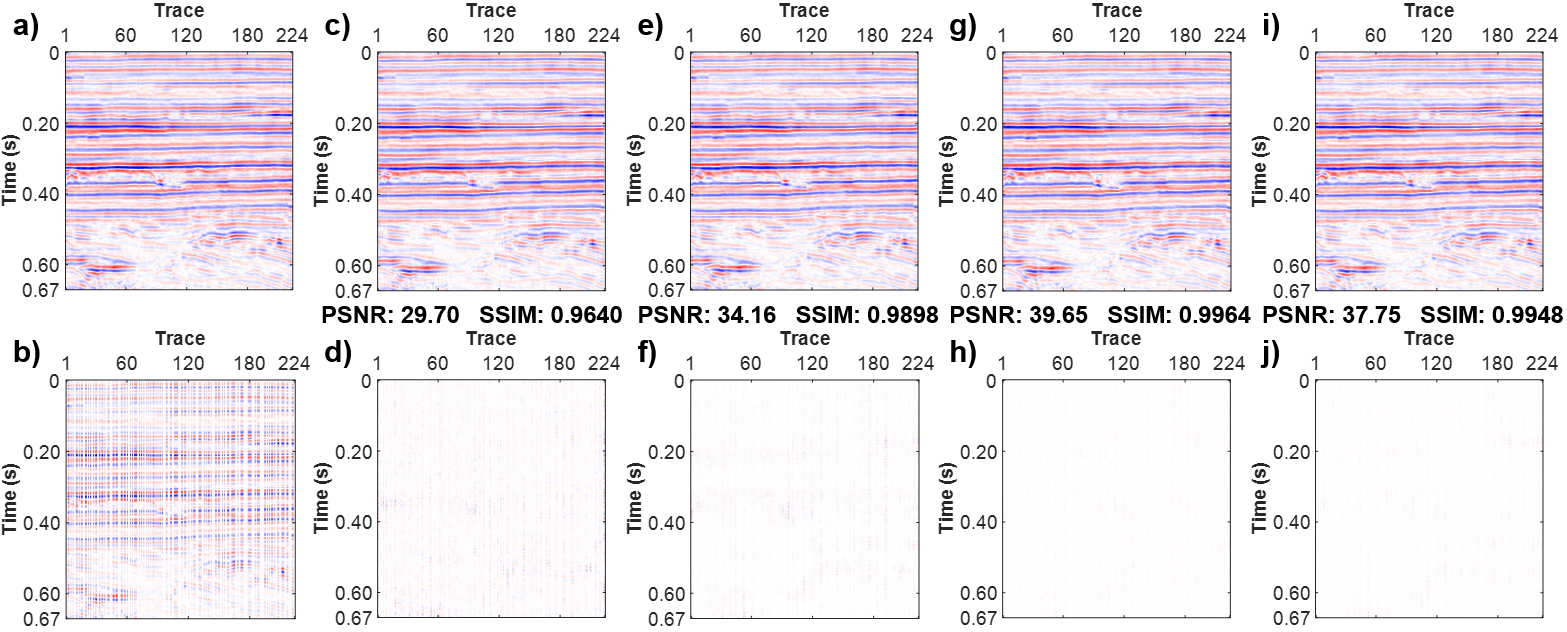}
\caption{
Comparisons of the tenth line extracted from Fig. \ref{fig13}. a) Complete data, (b) 50\% randomly sampled data, (c, e, g, and i) interpolated results of T-2D, T-3D, and T-2.5D, respectively; and (d, f, h, and j) the corresponding residual images.
}\label{fig14}
\end{figure}

Similarly, the tenth line is illustrated in Fig. \ref{fig14} as a 2D view. The numerical results in Fig. \ref{fig14}i and little leakage in Fig. \ref{fig14}j indicate the great interpolation ability of T-2.5D.

\begin{figure}
\centering
\includegraphics[width=1\textwidth]{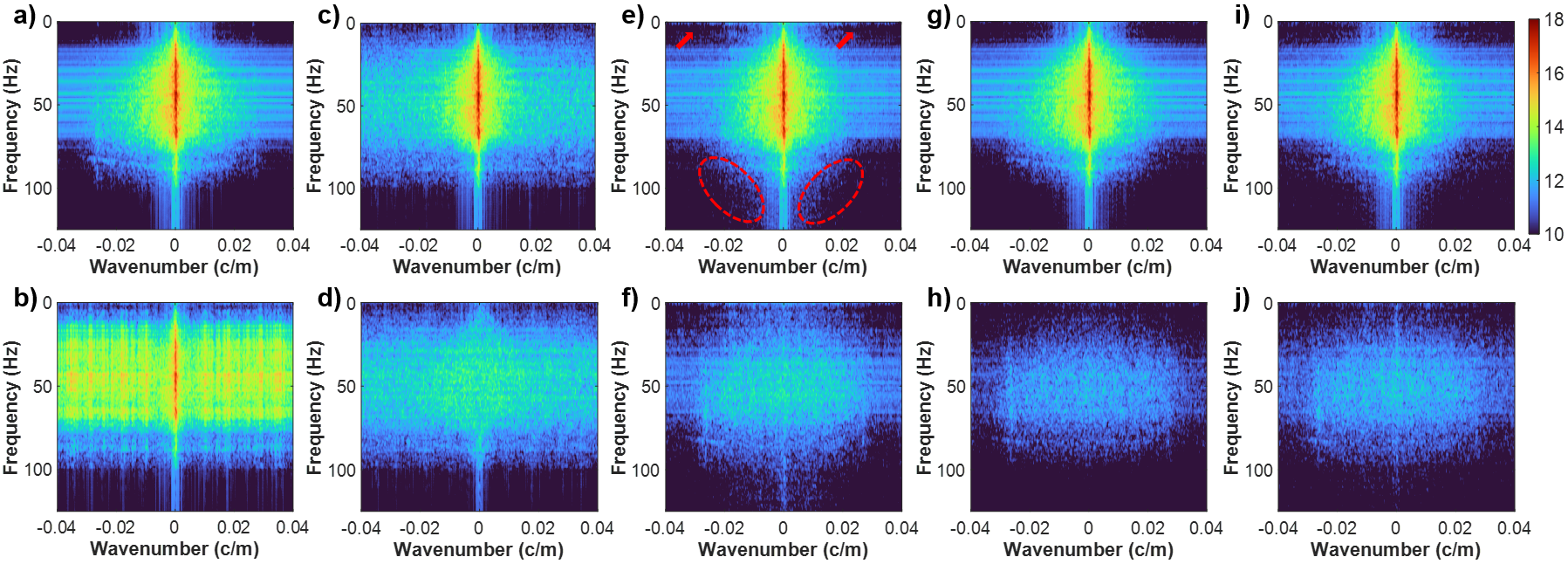}
\caption{
(a and b) f-k spectra of the complete data (Fig. \ref{fig14}a) and 50\% randomly sampled data (Fig. \ref{fig14}b), (c, e, g, and i) f-k spectra of the interpolated results by POCS, T-2D, T-3D and T2.5D (Fig. \ref{fig14}c, \ref{fig14}e, \ref{fig14}g, and \ref{fig14}i), respectively; and (d, f, h, and j) f-k spectra of the corresponding residual images (Fig. \ref{fig14}d, \ref{fig14}f, \ref{fig14}h and \ref{fig14}j).
}\label{fig15}
\end{figure}

Fig. \ref{fig15} displays the f-k spectra of the complete data (Fig. \ref{fig14}a), 50\% randomly sampled data (Fig. \ref{fig14}b), the interpolated results in Fig. \ref{fig14}c, \ref{fig14}e, \ref{fig14}g, and \ref{fig14}i, respectively, and the corresponding residual images (Fig. \ref{fig14}d, \ref{fig14}f, \ref{fig14}h, and \ref{fig14}j). As shown in Fig. \ref{fig15}c, strong residual interference exists in the result of POCS. T-2D still produces more residual interference than the other two methods do, as indicated by the red arrows and ovals in Fig. \ref{fig14}e. In comparison, the recovered spectrum in Fig. \ref{fig14}i and the energy leakage in Fig. \ref{fig14}j demonstrate the strong interpolation capability of T-2.5D.

\subsubsection{Analysis of the computational cost}\label{sec3.4.3}
Similar to Kerry dataset, we also provide comparisons of computational cost in Table \ref{tab4}. Due to the same settings of dataset preparation and training, the values on the Parihaka dataset are very similar to that of Kerry dataset. The T-2.5D has achieved great performance at very low computational cost, offering a good trade-off between the effectiveness and efficiency.
\begin{table}
\caption{Computational costs of T-2D, T-3D, and T-2.5D on the Parihaka dataset.}\label{tab4}%
\begin{tabular}{@{}lcc@{}}
\toprule
Method & Peak memory usage (MB) & Training time (h)  \\
\midrule
T-2D            & 776           & 0.56    \\
T-3D            & 56852         & 21.03    \\
T-2.5D & 4636 & 0.56 (Stage 1) + 0.65 (Stage 2)    \\
\bottomrule
\end{tabular}
\end{table}

\subsection{Generalization ability}\label{sec3.5}
\begin{figure}
\centering
\includegraphics[width=1\textwidth]{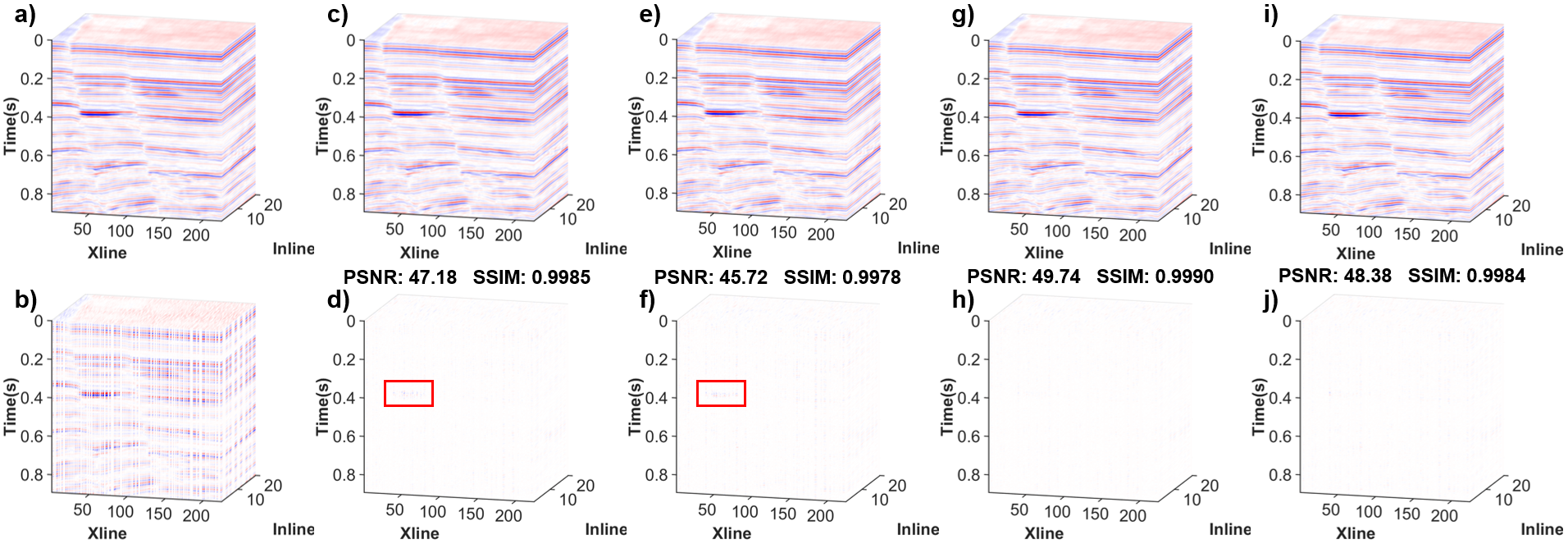}
\caption{
Interpolation of randomly sampled data on Opunake dataset. (a) Complete data, (b) 50\% randomly sampled data, (c and e) interpolated results of T-3D and T-2.5D before applying TL, respectively; (g and i) interpolated results of T-3D and T-2.5D after applying TL, respectively; (d, f, h, and j) corresponding residual images of (c), (e), (g) and (i), respectively.
}\label{fig16}
\end{figure}

Generalization is an important ability that allows a DL model to effectively process variant datasets that are not used during training \citep{zhang2021understanding}. To further explore the generalization capability, the Opunake dataset, which is from another 3D survey in New Zealand \citep{Opunake3D}, is utilized for further TL of T-2.5D and T-3D. The sampling interval of Opunake dataset is 0.004 s. The dataset preparation of the Opunake dataset is similar to that of the Kerry and Parihaka datasets. A 256×650×186 volume is extracted from the full Opunake dataset, and the first 166 lines are for training/validation and the last 20 lines are for testing. Here, TL for T-2.5D and T-3D is performed in a full fine-tuning manner. The comparisons of the interpolated results before and after the TL are presented in Fig. \ref{fig16}. As indicated by the red boxed in Fig. \ref{fig16}d and \ref{fig16}f, evident signal leakage can be observed in the residual images before applying TL. This phenomenon is significantly alleviated after applying TL, as shown in Fig \ref{fig16}g–\ref{fig16}j.

\begin{figure}
\centering
\includegraphics[width=1\textwidth]{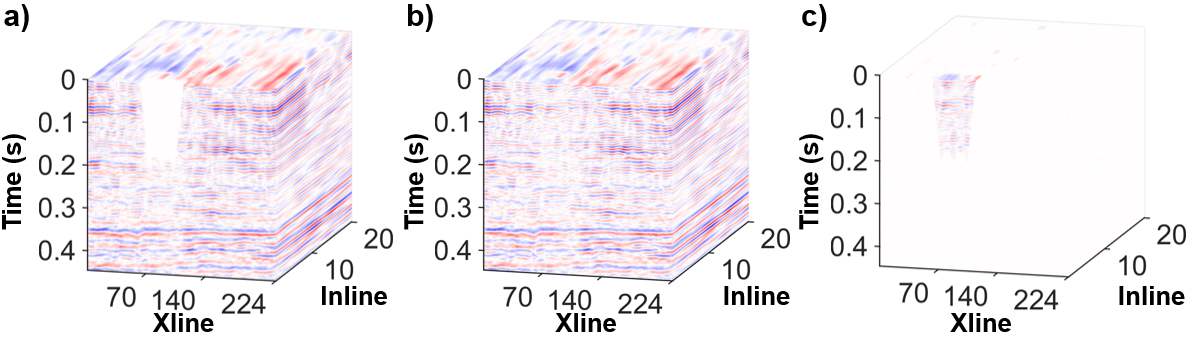}
\caption{
Interpolation of real missing data on Stratton dataset. (a) Incomplete input, (b) interpolated result of T-2.5D, and (c) the corresponding residual image.
}\label{fig17}
\end{figure}

To further evaluate the effectiveness on real missing data instead of that created by randomly removing traces, we extract a 3D volume from Stratton dataset, which is obtained from a land 3D survey in Texas, USA \citep{Stratton3D}. The 3D volume is tested using T-2.5D model with both stages trained on Kerry dataset, and without any additional fine-tuning on the target survey. As shown in Fig. \ref{fig17}a, missing data are mainly distributed in time 0–0.19 s, Xline 57–108 and Inline 1–2 within the 3D volume. The interpolated result in Fig. \ref{fig17}b shows a good continuity of the reconstructed events. We also provide the residual image in Fig. \ref{fig17}c, which is defined as the difference between the interpolated result in Fig. \ref{fig17}b and the incomplete input in Fig. \ref{fig17}a. Since the true signals at the missing regions are unknown, this example serves as a qualitative continuity evidence of the reconstructed events.

In summary, T-2.5D has great potential for effectively interpolating missing traces in 3D seismic data from different survey areas, exhibiting great generalization ability.

\section{Discussions}\label{sec4}
\subsection{Testing in the Regularly Missing Case}\label{sec4.1}
All the interpolated results in the previous section are dealing with irregularly missing traces. In fact, T-2.5D can also be leveraged to handle regular gaps to increase the acquisition density. To validate this, we further evaluate the performance of T-2.5D on the Kerry and Parihaka with regularly missing traces. We conduct a series of additional experiments on the test data of the two datasets under two missing rates, which are 67\% (removing two out of every three traces) and 80\% (removing four out of every five traces). The quantitative results are shown in Table \ref{tab5}, where T3D exhibits severe performance degradation under extremely high missing rates. Meanwhile, T2D and T-2.5D maintain comparable interpolation capabilities. One possible reason for the limited advantage of T-2.5D in this scenario is the relatively small training set sizes and the moderate random masking ratio adopted in this work (40\%–60\%).
\begin{table}
\caption{Comparisons on the Kerry and Parihaka datasets under two regularly missing rates.}\label{tab5}%
\begin{tabular}{@{\extracolsep\fill}lcccc}
\toprule
                & \multicolumn{4}{c}{PSNR (dB)}
\\\cmidrule{2-5}
Method & Kerry 67\% & Kerry 80\% & Parihaka 67\% & Parihaka 80\% \\
\midrule
T-2D   & 32.81 & 26.89 & 34.31 & 27.69 \\
T-3D   & 32.92 & 21.68 & 34.90 & 25.36 \\
T-2.5D & 33.01 & 26.71 & 34.77 & 27.78 \\
\bottomrule
\end{tabular}
\end{table}

\begin{figure}
\centering
\includegraphics[width=1\textwidth]{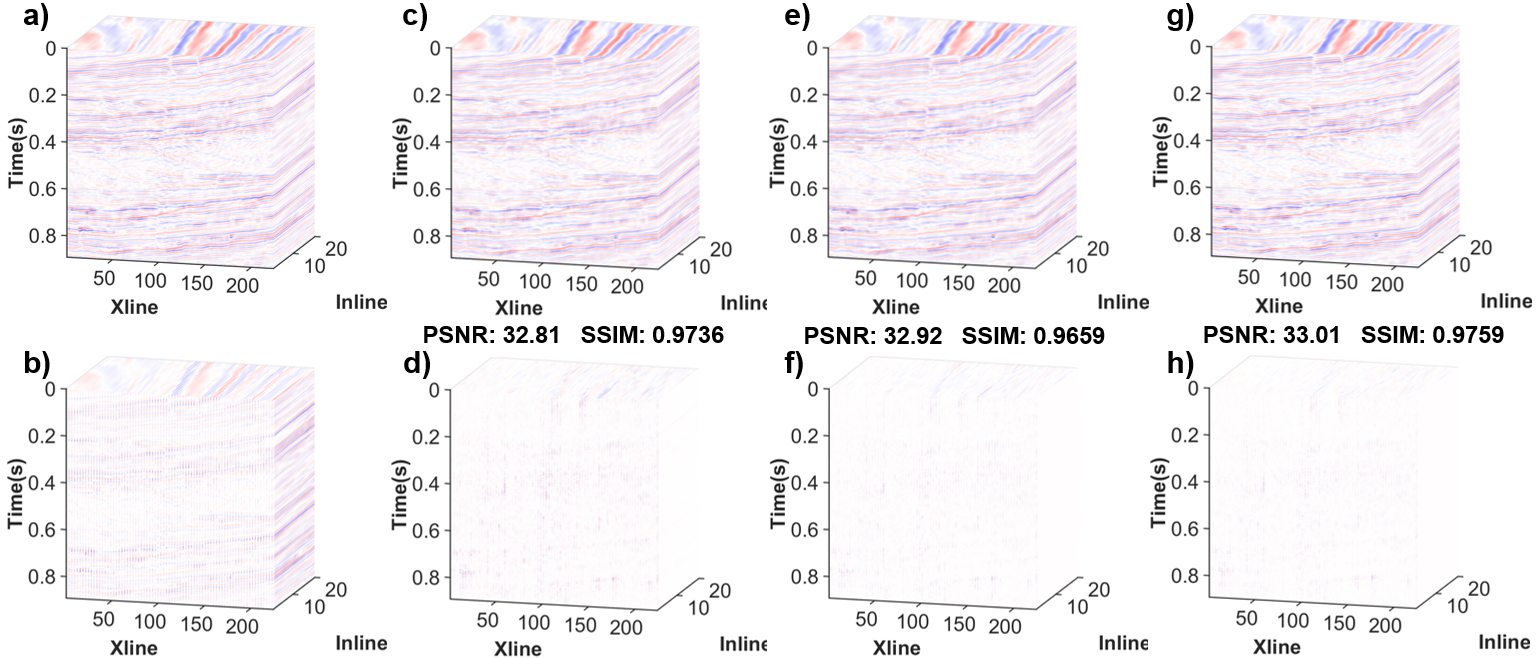}
\caption{
Interpolation of regularly sampled data on Kerry dataset. (a) Complete data, (b) 67\% regularly sampled data, (c, e, and g) interpolated results of T-2D, T-3D, and T-2.5D, respectively; (d, f, and h) the corresponding residual images.
}\label{fig18}
\end{figure}

The interpolated results under 67\% regularly missing rate are displayed in Fig. \ref{fig18}. where the interpolation result of T-2.5D shows superior performance to T-2D. Furthermore, it has achieved even better numerical results than T-3D does under such a high missing rate. This further validates the effectiveness of T-2.5D for seismic interpolation under sparse acquisition scenarios.

\subsection{Comparison of Parameters}\label{sec4.2}
The number of parameters in a network model is also an important factor for DL-based methods. Models with larger capacity or parameters can represent more complex functions, although they generally require greater memory and computational resources \citep{shazeer2017outrageously, baldi2019capacity}. We further compare the parameters of the three networks. Notably, the total parameters of T-2.5D consist of 34,048 pretrained parameters transferred from Stage 1 and 1,319,121 parameters optimized in Stage 2. As shown in Table \ref{tab6}, the parameters surge from 2D to 3D. Although T-2D has the fewest parameters, its interpolation performance is also inferior to the other two networks. Although the total parameters of T-2.5D are only 75.34\% of those of T-3D, T-2.5D can achieve performance comparable to T-3D, or even surpassing it in certain cases. This finding further supports the lightweight and efficient nature of T-2.5D.
\begin{table}
\caption{Parameters in different networks. }\label{tab6}%
\begin{tabular}{@{}lc@{}}
\toprule
Method & Parameters  \\
\midrule
T-2D            & 212289    \\
T-3D            & 1795953    \\
T-2.5D & 34048 (inherited from Stage 1) + 1319121 (Stage 2)    \\
\bottomrule
\end{tabular}
\end{table}

\subsection{Comparisons of inference time}\label{sec4.3}
In DL-based seismic data pre-processing, the inference time also plays an important role due to the need for processing massive datasets. Hence, we have recorded the inference time of T-2D, T-3D and T-2.5D in randomly missing cases. During inference, we used a batch size of 1. The timing started when the data were first fed into the model in the sliding-window manner and ended when the complete 3D volume was output, which means T-2D may need time to reassemble 2D patches to an entire 3D volume. As shown in Table \ref{tab7}, the inference time of T-2.5D is even less than that of T-2D. This is because T-2D requires additional processing time for splitting and reassembling the 3D volumes, which is not needed in T-2.5D. Meanwhile, it only requires less than 1/9 the inference time of T-3D. These comparisons demonstrate that T-2.5D is capable of interpolating seismic data rapidly with high quality when faced with large-scale 3D seismic datasets.
\begin{table}
\caption{Inference time of T-2D, T-3D, and T-2.5D.}\label{tab7}%
\begin{tabular}{@{\extracolsep\fill}lcccc}
\toprule
 & \multicolumn{4}{c}{Inference time (s)}
\\\cmidrule{2-5}
Method & Kerry dataset & Parihaka dataset & Opunake Dataset & Stratton Dataset   \\
\midrule
T-2D        & 9.47  & 9.39  & N/A    & N/A\\
T-3D        & 59.04 & 58.98 & 58.04  & 58.95 \\
T-2.5D      & 6.69  & 6.68  & 6.83  & 6.71 \\
\bottomrule
\end{tabular}
\end{table}

The substantial acceleration of T-2.5D can be explained from the computational complexity of the self-attention mechanism. In Transformer architectures, the computational complexity of MSA grows quadratically with the sequence length.  It is denoted as $O(N^2)$, where $N$ is the sequence length. For T-3D, the processing time of MSA on a 3D volume of shape T×X×I is $c(TXI)^2$, where $c$ is a constant used to convert this complexity to processing time, and T, X, and I are the three dimensions introduced in Section \ref{sec2.2.2}. Therefore, the inference time of T-3D can be expressed as
\begin{equation}\label{eq15}
    t_{3D}=c(TXI)^2+t^{3D}_{res},
\end{equation}
where $t^{3D}_{res}$ denotes the processing time of the remaining network components. Since the T-2.5D model performs MSA on every 2D slices as described in Section \ref{sec2.2.2}, the processing time of MSA becomes $cI(TX)^2$. On this basis, the inference time of T-2.5D can be calculated by
\begin{equation}\label{eq16}
    t_{2.5D}=cI(TX)^2+t^{2.5D}_{res},
\end{equation}
where $t^{2.5D}_{res}$ represents the processing time of the remaining network components. Theoretically, for T-2.5D, the computational complexity of the MSA is reduced to 1/I (where I=16 in our implementation) of that of T-3D. Therefore, considering the additional computational cost introduced by the SDAs together with the remaining network components, the observed reduction of the total inference time to approximately 1/9 of that of T-3D is reasonable.

\subsection{Interpolation on complex structural data}\label{sec4.4}
\begin{figure}
\centering
\includegraphics[width=1\textwidth]{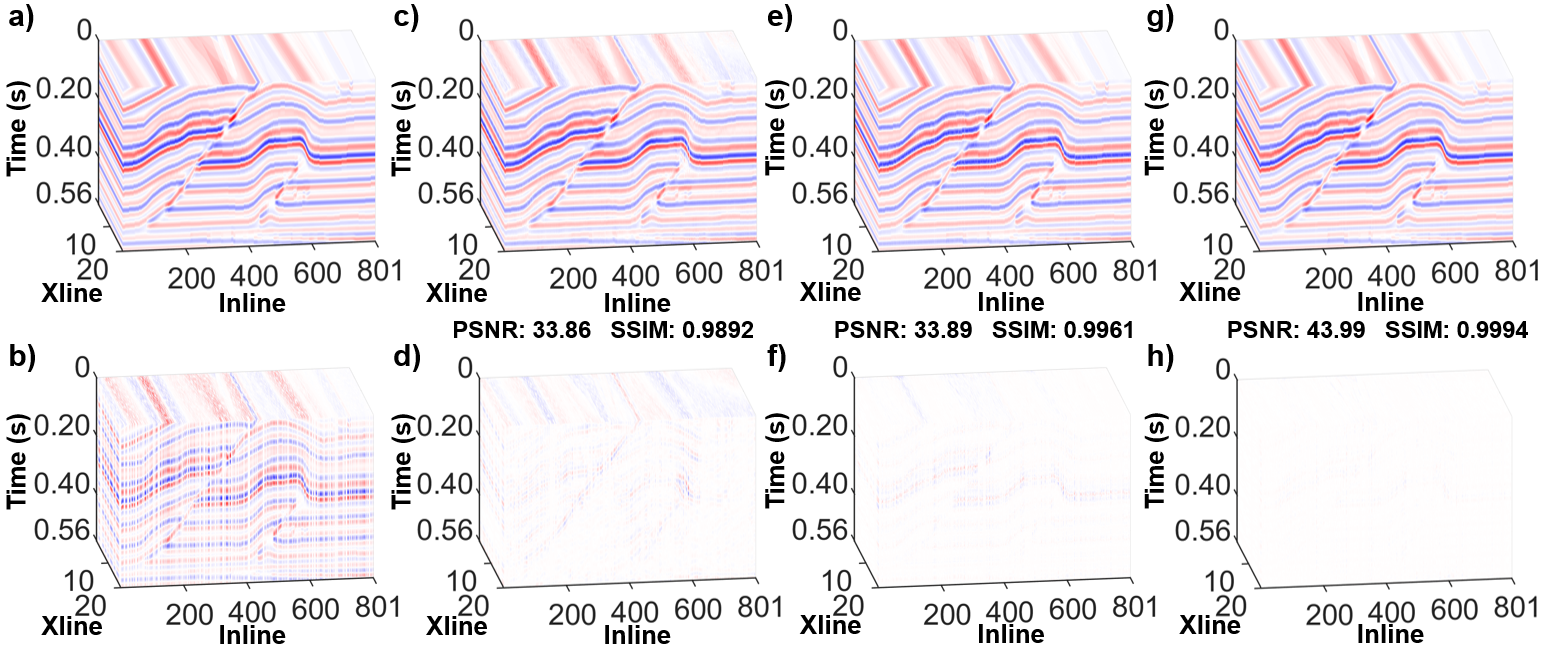}
\caption{
Interpolation of irregularly sampled data on Overthrust dataset. (a) Complete data, (b) 50\% irregularly sampled data, (c, e, and g) interpolated results of POCS, T-2.5D-K, and T-2.5D-K-O, respectively; (d, f, and h) the corresponding residual images.
}\label{fig19}
\end{figure}

In this subsection, we discuss the interpolation capability of T-2.5D on complex structural data. A synthetic 3D volume with a shape of 207×801×801 is generated using SEG/EAGE Overthrust model \citep{Aminzadeh1997}. The Overthrust model is a geologically complex synthetic model containing multiple geological features, and the convergence of major faults and pronounced three-dimensional structural variations introduce substantial challenges for seismic interpolation. We extract time samples 30–170, xline indices 601–680, and inline indices 1–801 for training and validation, and generate 2000 and 500 samples, respectively. Meanwhile, time samples 30–170, xline indices 681–700, and inline indices 1–801 are used for testing. We load the existing weights trained on Kerry dataset into the 2D Transformer encoders, and optimize 3D SDAs using the samples above. This trained model is denoted as T-2.5D-K-O. The aforementioned POCS method, as well as T-2.5D with both stages trained on Kerry dataset (denoted as T-2.5D-K) are used as comparison methods. As shown in Fig. \ref{fig19}, among the three methods, POCS produces the most leakage. The result in Fig. \ref{fig19}e suffers from the generalization ability of T-2.5D-K across synthetic and field seismic datasets. In comparison, T-2.5D-K-O reconstructs the most continuous data with highest numerical results in Fig. \ref{fig19}g, and produces the least signal leakage in Fig. \ref{fig19}h. In summary, T-2.5D is a flexible two-stage interpolation method, and is capable of interpolating data with high structural complexity.

\subsection{Does mixed-survey training improve interpolation performance?}\label{sec4.5}
One remaining issue in this study is that the model needs to be trained separately for each seismic survey, making the experiments somewhat repetitive. Would it not be more advantageous to train a single model using a dataset comprising multiple seismic surveys? Motivated by this question, we conducted the following experiment. We additionally investigated mixed-survey training using T-2D. Specifically, 2,000 training patches and 500 validation patches were randomly selected from each of the Kerry, Parihaka, and Opunake datasets, maintaining a total of 6,000 training and 1,500 validation patches. The resulting model was evaluated on the Kerry and Parihaka datasets, and the quantitative comparisons between single-dataset-trained model and mixed-dataset-trained model are summarized in Table \ref{tab8}. Compared with models trained on individual datasets, the mixed-data model even exhibits lower numerical results, indicating weaker amplitude reconstruction. These results suggest that mixed-survey training did not yield performance improvements under the current training configurations.
\begin{table}
\caption{Comparisons between single-dataset-trained model and mixed-dataset-trained model.}\label{tab8}%
\begin{tabular}{@{\extracolsep\fill}lcccc}
\toprule
                & \multicolumn{2}{c}{Kerry test data} & \multicolumn{2}{c}{Parihaka test data}
\\\cmidrule{2-5}
Models                                  & PSNR & SSIM & PSNR & SSIM   \\
\midrule
T-2D trained with Kerry dateset         & 35.23 & 0.9837    & N/A    & N/A\\
T-2D trained with Parihaka dataset      & N/A   & N/A       & 36.51  & 0.9902 \\
T-2D trained with mixed dataset         & 35.09 & 0.9827    & 35.07  & 0.9903 \\
\bottomrule
\end{tabular}
\end{table}

\section{Conclusion}
In this paper, we propose a Transformer-based network, named T-2.5D, to accomplish 3D seismic data interpolation tasks. Specifically, a cross-dimensional TL training strategy is utilized to optimize T-2.5D, so as to reduce the computational burdens when interpolating 3D seismic data with Transformer. This strategy consists of two stages. The first stage is the 2D pre-training stage. It generates a pre-trained model by optimizing the Transformer encoders in T-2.5D using a large amount of 2D data patches. The second stage is the 3D fine-tuning stage, in which the 3D SDAs are fine-tuned using a small number of 3D data volumes. In this stage, the SDAs, each placed before a Transformer encoder, can learn abundant spatial correlation information across seismic lines. Extensive experiments demonstrate that the T-2.5D with this cross-dimensional strategy exhibits comparable or even better performance to T-3D, while requiring only small amounts of memory and training time. In summary, the proposed T-2.5D is qualified to replace time-consuming, memory-intensive full 3D training, so as to significantly improve the efficiency of 3D interpolation using Transformer.

\begin{acknowledgments}
This work was jointly supported by the Deep Earth Probe and Mineral Resources Exploration–National Science and Technology Major Project under Grant 2025ZD1007603, and National Natural Science Foundation of China under Grant 42574169.
\end{acknowledgments}

\section{Data availability statement}
The data used in this paper is available at \href{https://wiki.seg.org/wiki/Open\_data}{https://wiki.seg.org/wiki/Open\_data}. The supplementary materials supporting this study are publicly available in our \href{https://github.com/YChonsin/T-2.5D_Project}{GitHub repository}. The source code is currently being prepared for public release and will be made available in the same repository upon publication.

\bibliographystyle{gji}
\bibliography{ref}

%\appendix

%Table~\ref{authors} is a list of design macros which are unique to GJI. The
%list displays each macro's name and description.
%
%\section{For editors}
%
%The additional features shown in Table~\ref{editors} may be used for
%production purposes.

%\bsp % ``This paper has been produced using the Blackwell
%     %   Publishing GJI \LaTeXe\ class file.''

\label{lastpage}

\end{document}